\documentclass[12pts]{article}
\usepackage{lineno}

\usepackage[dvips]{graphicx}
\usepackage[toc,page]{appendix}
\usepackage{graphicx}
\usepackage{amsmath}
\usepackage[varg]{txfonts}

\usepackage{longtable}

\usepackage[usenames]{color}
\usepackage{xspace}
\usepackage{siunitx}

\usepackage{natbib}
\begin{document}

\title{Global occurrence and chemical impact of stratospheric Blue Jets modeled with WACCM4}

\author{
  F. J. P\'erez-Invern\'on$^{1}$,
 F. J. Gordillo-V\'azquez$^{1}$, \\
A. K. Smith$^{2}$,
 E. Arnone$^{3,4}$,
H. Winkler$^{5}$.\\
\textit{$^{1}$Instituto de Astrof\'isica de Andaluc\'ia (IAA),} \\
   \textit{CSIC, PO Box 3004, 18080 Granada, Spain.}\\
\textit{$^{2}$Atmospheric Chemistry Observations and Modeling, } \\
   \textit{National Center for Atmospheric Research, Boulder, CO, USA.}\\
\textit{$^{3}$Istituto di Scienza dell'Atmosfera e del Clima - CNR, Bologna, Italy.}\\
\textit{$^{4}$Dipartimento de Fisica, 
 Universit\`a degli Studi di Torino, Italy.}\\
\textit{$^{5}$Institute of Environmental Physics, 
 University of Bremen, Germany.}\\
\footnote{Correspondence to: fjpi@iaa.es. 
Article published in Journal of Geophysical Research: Atmospheres.}
}
\date{}
\maketitle

\begin{abstract}
In this work we present the first parameterizations of the global occurrence rate and chemical influence of Blue Jets, a type of Transient Luminous Event (TLE) taking place in the stratospheric region above thunderclouds. These parameterizations are directly coupled with five different lightning parameterizations implemented in the Whole Atmosphere Community Climate Model (WACCM4). We have obtained a maximum Blue Jet global occurrence rate of about 0.9 BJ per minute. The geographical occurrence of Blue Jets is closely related to the chosen lightning parameterization.  Some previously developed local chemical models of Blue Jets predicted an important influence onto the stratospheric concentration of N$_2$O, NO$_x$ and O$_3$. We have used these results together with our global implementations of Blue Jets in WACCM4 to estimate their global chemical influence in the atmosphere. According to our results, Blue Jets can inject about 3.8 Tg N$_2$O-N yr$^{-1}$ and 0.07 Tg NO-N yr$^{-1}$ near the stratosphere, where N$_2$O-N and NO-N stand for the mass of nitrogen atoms in N$_2$O and NO molecules, respectively. These production rates of N$_2$O and NO$_x$ could have a direct impact on, for example, the acidity of rainwater or the greenhouse effect. We have found that Blue Jets could also slightly contribute to the depletion of stratospheric ozone. In particular, we have estimated that the maximum difference in the concentration of O$_3$ at 30 km of altitude between simulations with and without Blue Jets can be about -5 \% in Equatorial and Polar regions.

\end{abstract}

\section{Introduction}
\label{sec:intro}
The electric field produced as a consequence of the separation of electrical charges inside clouds is the origin of lightning in the troposphere. However, as originally proposed by \cite{Wilson1925/PPhSocLon} and later detected by \cite{Franz1990/Sci}, atmospheric electrical discharges can also take place in upper regions of the atmosphere. These types of electrical phenonema are known as Transient Luminous Events (TLEs).

The first detected TLE was a sprite \citep{Franz1990/Sci}, an upper atmospheric discharge formed by a complex structure of thousands of streamers and a diffuse non-streamer zone that can extend from 40~km up to 85~km of altitude \citep{Sentman1994/VIDEO, Lyons1994/GeoRL}. Other types of TLEs, known as halos and elves, can also be produced in the upper atmosphere at altitudes greater than 70~km and 80~km. Both halos and sprites can have a duration of several milliseconds \citep{Lyons2000/ETAGU, Barrington-Leigh2001/JGR,Wescott2001/JGR/1, Bering2002/AdSpR,Moudry2003/JASTP,Bering2004/AdSpR,Bering2004/GeoRL,Frey2007/GeoRL,Sentman2008/JGRD/1, Gordillo-Vazquez2008/JPhD, Luque2011/NatGe}, while elves have a duration of less than 1~ms \citep{Inan1991/GRL, Inan1997/GeoRL, Taranenko1993/GRL, Moudry2003/JASTP,  Kuo2007/JGRA, Marshall2010/JGRA/2, gordillo2016upper, van_der_Velde2016/GRL, perezmodeling, perezspectroscopic}.

In 1995, \cite{Wescott1995/GeoRL, Wescott1996/GeoRL/1} discovered the existence of upward propagating conical-shaped jets in the ranges of altitudes between 15~km and 25~km. Later in year 2002, \cite{Pasko2002/Natur} reported another type of upward propagating jets that reached the ionosphere.  These upward propagating discharges were later called Blue Jets (BJ) and Gigantic Jets (GJ), two types of TLEs that can propagate in the range of altitudes between 15~km and 40~km in the case of Blue Jets, and between 15~km and 90~km in the case of GJs.  The upper altitude reached by Blue Jets (about 40~km) corresponds to the level where propagation time equals the relaxation timescale of about 0.2~s \citep{Sukhorukov1996/GeoRL/1}. Blue Jets and Gigantic Jets are different events triggered right above the cloud layer \citep{Pasko2002/Natur, van_der_Velde2010/JGR, Pasko2012/SSR, Chanrion2017/GRL}.  According to some evidences \citep{Krehbiel2008/NatGe, Riousset2010/JGRA, Pasko2012/SSR}, Gigantic Jets could be initiated as a cloud lightning discharge propagating upward, while Blue Jets are triggered as a consequence of the electrical breakdown produced between the storm upper charge layer and the screening charge attracted to the cloud top \citep{Krehbiel2008/NatGe, Riousset2010/JGRA, Pasko2012/SSR}.

Since their discovery in 1989, TLEs have been observed from planes, balloons, ground-based detectors and space-based instrumentation. Several campaigns have recorded the spectra of sprites \citep{Hampton1996/GeoRL, Kanmae2007/GeoRL, Passas2016/APO, Gordillo-Vazquez2018/JGR}. Some space-based missions, such as the Space Shuttle \citep{Boeck1992/GRL}, the Imager of Sprites and Upper Atmospheric Lightning (ISUAL) of the National Space Organization of Taiwan (NSPO) \citep{Chern2003/JASTP, Chen2008/JGRA, hsu2017/TAOC} and the Global Lightning and sprIte MeasurementS (GLIMS) of the Japan Aerospace Exploration Agency (JAXA) \citep{sato2015overview,Adachi2016/JASTP} have reported TLE observations from space. Last April 2, 2018 the Atmosphere-Space Interactions Monitor (ASIM) \citep{Neubert2006/ILWS} of the European Space Agency (ESA) was successfully launched. ASIM is equipped with the Modular Multi-Imaging Assembly (MMIA), devoted to the study of TLEs from space. In addition, the Tool for the Analysis of RAdiations from lightNIng and Sprites (TARANIS) \citep{Blanc2007/AdSpR} of the Centre National d'\'Etudes Spatiales (CNES), will also be devoted to the observation of these events after its expected launch in 2019 or 2020.

Several authors have investigated the local chemical impact of TLEs \citep{Gordillo-Vazquez2008/JPhD, Sentman2008/JGRD/1, Gordillo-Vazquez2009/PSST, Gordillo-Vazquez2010/JGRA, Pasko2012/SSR, Parra-Rojas/JGR, Parra-Rojas/JGR2015, Winkler2015/JASTP, PerezInvernon2016/GRL, Hoder2016/IOP, perezmodeling}. Recently, \cite{Winkler2015/JASTP} developed a local chemical model of Blue Jets obtaining an important local enhancement of NO$_x$, N$_2$O and O. The global chemical influence of TLEs has been investigated by previous studies. According to previous local models of halos and elves \citep{perezmodeling}, their global chemical impact would be negligible. \cite{Arnone2014/JGR} estimated the global production of NO$_x$ by sprites using the Whole Atmosphere Community Climate Model version 4 (WACCM4). \cite{Arnone2014/JGR} found that a perturbation in the tropical concentration of nitrogen oxide by sprites could lie between 0.015~ppbv and 0.15~ppbv. These quantities correspond to a perturbation of the background concentration of NO$_x$ between less than 1 \% and up to 20 \% at different altitudes. Some observational studies have attempted to measure sprite-NO$_x$ through satellite observations \citep{Arnone2008/GeoRL, Rodger2008/GRL, Arnone2009/PSST, arnone2016chimtea}. However, according to these studies sprite-NO$_x$ is at the edge of current detectability.
The predicted significant local chemical influence \citep{Winkler2015/JASTP} suggest that Blue Jets could have a non-negligible influence in the chemistry of the atmosphere.

In this work, we have developed the first global parameterization of Blue Jets. We have used the WACCM4 model in order to study the global occurrence rate of Blue Jets and their global chemical impact by developing three different Blue Jet parameterizations. WACCM4 includes a lightning parameterization developed by \cite{Price1992/JGR} based on the cloud top height (CTH). Here we also use other lightning parameterizations based on, respectively, the amount of convective precipitation (CP) \citep{Allenp2002/JGR}; the upward mass flux (MFLUX) \citep{Allenp2002/JGR}; the precipitation rate and the Convective Available Potential Energy (CPCAPE) \citep{Romps2014/SCI}; and on the upward cloud ice flux (ICEFLUX) \citep{Finney2014/ACP}.The combined use of lightning and Blue Jet parameterizations allow us to predict the geographical and seasonal chemical impact of Blue Jets.


\subsection{Physics and chemistry of Blue Jets} 
\label{sec:bjstatistics}

\begin{figure}
\includegraphics[width=0.6\columnwidth]{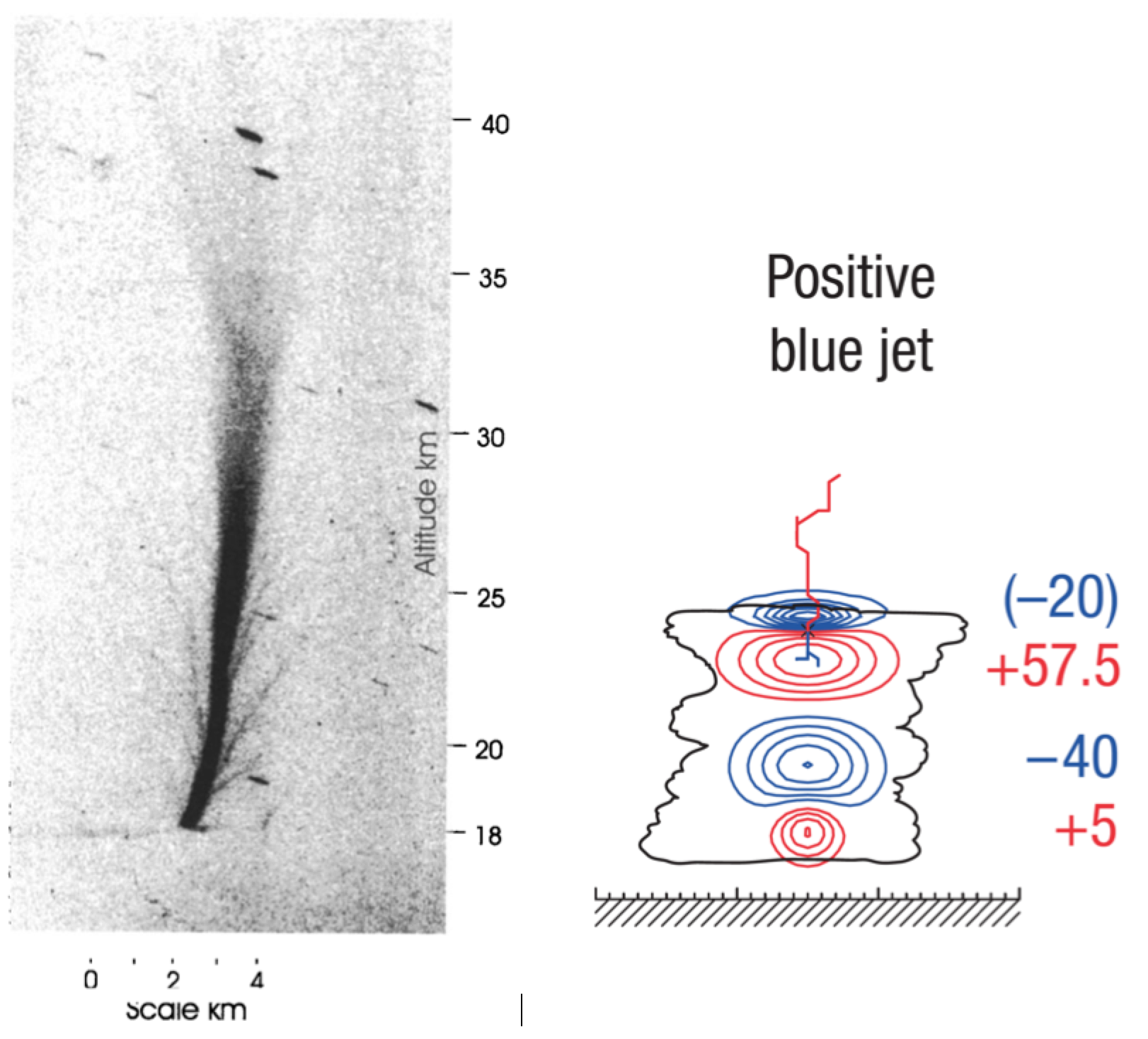}
\footnotesize
\caption{\label{fig:BJ_wescott}
Left panel: Inverted black and white photography of a Blue Jet. The spatial scales of the leader and streamer regions can be appreciated. Image adapted from \cite{Wescott2001/JGR}. Right panel: Blue Jet simulated by \cite{Krehbiel2008/NatGe} illustrating the charge structure of clouds. Blue and red lines correspond to positive and negative charges, respectively. Image adapted from \cite{Krehbiel2008/NatGe}.
\normalsize
}
\end{figure}

Blue Jets are formed by a leader channel surrounded by a large number of streamers. The leader is a highly conductive plasma channel that can heat the air up to thousands of Kelvin. The first interpretations of Blue Jets by a streamer corona of a leader were made by \cite{Sukhorukov1998/JASTP} and \cite{Petrov1999/JTePh}. \cite{Raizer2006/GeoRL, Raizer2007/JASTP} proposed the development of Blue Jets as a bi-leader channel that propagates upward from the streamer zone of a positive leader. According to \cite{Raizer2006/GeoRL, Raizer2007/JASTP}, this leader can transfer the energy contained in the clouds to upper regions of the atmosphere, where the low density allows the development of a streamer corona. Figure~\ref{fig:BJ_wescott} shows a photography of a real Blue Jet from\cite{Wescott2001/JGR} and the structure of charges in clouds that trigger the inception of Blue Jets simulated by \cite{Krehbiel2008/NatGe}.
 \cite{Krehbiel2008/NatGe} developed a model based on quasielectrostatic fields, formed as an imbalance of the electric charge in the cloud tops, to predict Blue Jet and Gigantic Jet inception. After lightning occurs, a charged layer can remain near the storm top layer creating a local electric field. \cite{Krehbiel2008/NatGe} found that conventional electric breakdown near this charged layer could trigger an upward propagating leader, forming a Blue Jet. \cite{Riousset2010/JGRA} upgraded the model proposed by \cite{Krehbiel2008/NatGe}, confirming the obtained results. Observations by \cite{Lu2011/GRL} supported some of the predictions by \cite{Krehbiel2008/NatGe} and \cite{Riousset2010/JGRA}. As hypothesized by \cite{Krehbiel2008/NatGe}, there would exist a competition between intra-cloud discharges and Blue Jets in the process of discharging the cloud. The result of this competition would depend on the capability of the convective fluxes to mix the oppositely charged layers located in the cloud top before the inception of a Blue Jet.

\cite{Chanrion2017/GRL} have recently described the observation of a Blue Jet from the International Space Station (ISS). The top height of the thundercloud that initiated the Blue Jet reached the tropopause, an atmospheric region where convection is weak. The reported Blue Jet was preceded in 1.16~s by a strong negative CG lightning with a peak current of -167.5~kA. This CG discharge could possibly be the parent lightning of the Blue Jet.
A Blue Jet would then be formed by an upward propagating leader traveling to upper regions of the atmosphere with lower pressure, reaching its maximum altitude between 30~km and 40~km \citep{van_der_Velde2010/JGR, Pasko2012/SSR, daSilva2013/GRL, Milikh2014/JGR, Chanrion2017/GRL}. Streamers could then emerge from the leader as it passes through the low pressure regions of the atmosphere \citep{Raizer2007/JASTP}.

\cite{Mishin1997/GeoRL} and \cite{smirnova2003/IJGA} developed the first models to estimate the local chemical impact of Blue Jets. However, these models do not include the latest results on the electrodynamical mechanisms of Blue Jets \citep{Raizer2007/JASTP, Krehbiel2008/NatGe, Riousset2010/JGRA}. \cite{Winkler2015/JASTP} developed the most detailed model to date to study the local chemical impact of a Blue Jet including 88~species interacting through more than 1000~reactions. They used their model to estimate the local chemical impact of the leader and streamers of a Blue Jet at several altitudes.
According to their estimations, the high-temperature reactions taking place in the Blue Jet leader can enhance by several orders of magnitude the local background concentrations of stratospheric N$_2$O and NO (due to the high temperature reactions collected in Table~3 of \citep{Winkler2015/JASTP}) and produce a significant depletion of ozone \citep{Winkler2015/JASTP}. In addition, the high electric field in the streamer phase would produce an enhancement in the concentration of N$_2$O by the chemical reactions

\begin{linenomath*}
\begin{equation}
e + N_2 \rightarrow e + N_2(A^3\Sigma_u^+) \label{reactionN2A}
\end{equation}
\end{linenomath*}
and
\begin{linenomath*}
\begin{equation}
N_2(A^3\Sigma_u^+) + O_2 \rightarrow N_2O + O. \label{reactionN2A}
\end{equation}
\end{linenomath*}

The injection of NO$_x$ into the stratosphere could also influence the concentration of other species. According to investigations about the chemical influence of lightning-produced NO in the atmosphere, atmospheric electricity phenomena can also contribute to the concentration of OH, HO$_2$ and CO \citep{rohrer2006strong, murray2013interannual, siingh2015lightning}. In particular, NO interacts with HO$_2$ producing OH. The production of OH molecules can influence the acidity of rainwater \citep{seinfeld2016atmospheric}, as they can react with NO$_2$ molecules producing HNO$_3$ following the chemical reaction NO$_2$ + OH + M $\rightarrow$ HNO$_3$ + M \citep{labrador2005effects}. The formation of OH contributes to the loss of CO by the process  CO + OH $\rightarrow$ HO$_2$ + C \citep{murray2013interannual}. OH molecules can also contribute to the oxidation of SO$_2$, leading to the production of H$_2$SO$_4$. In addition, NO$_2$ molecules contribute to the production of N$_2$O$_5$. The oxidation of N$_2$O$_5$ followed by a heterogeneous hydrolysis reaction on aerosol particles contributes to the enhancement of HNO$_3$.


\subsection{Global budgets of N$_2$O and NO$_x$ and their relation with atmospheric electricity} 
\label{sec:budjet}

According to \cite{Winkler2015/JASTP}, Blue Jets could inject an important amount of nitrous oxide (N$_2$O), nitric oxide (NO) and nitrogen dioxide NO$_2$ at stratospheric altitudes. These gases play important roles in the chemical balance of stratospheric ozone. In addition, N$_2$O is one of the most important greenhouse gases. 

Natural N$_2$O sources are estimated to inject about 10.2~Tg~N$_2$O-N~yr$^{-1}$ in the atmosphere, while anthropogenic sources could produce around 6.3~Tg N$_2$O-N~yr$^{-1}$ \citep{davidson2009/nat, Prather2015/JGR}, where 1~Tg = 10$^{12}$~g and N$_2$O-N stands for the mass of nitrogen atoms in N$_2$O molecules  \citep{davidson2009/nat}. The major natural and anthropogenic sources of N$_2$O are basically due to nitrification and denitrification produced by microbes at ground level \citep{davidson2009/nat}. However, \cite{Sheese2015/GRL} have recently proposed an atmospheric source of N$_2$O based on observations from the satellite instrument ``Atmospheric Chemistry Experiment-Fourier Transform Spectrometer" (ACE-FTS) consisting of the chemical reaction described in equation~(\ref{reactionN2A}) \citep{arnone2012stratosphere, Sheese2015/GRL}. N$_2$O is the major source of NO in the stratosphere. 90 \% of the stratospheric destruction of N$_2$O is by photolysis (N$_2$O + h$\nu$ $\rightarrow$ N$_2$ + O) and 10 \% is by reaction with O($^1$D) producing NO, N$_2$ and O$_2$ molecules \citep{seinfeld2016atmospheric}. 


\cite{Plieninger2016/ACP} compared global-average vertical profiles of N$_2$O obtained by different instruments. In particular, \cite{Plieninger2016/ACP} showed the vertical stratospheric concentration of N$_2$O  obtained by the ``Michelson Interferometer for Passive Atmospheric Sounding" (MIPAS), the ``Atmospheric Chemistry Experiment-Fourier Transform Spectrometer (ACE-FTS), the ``Microwave Limb Sounder onboard Aura" (Aura-MLS) and the ``Sub-Milimetre Radiometer onboard Odin" (Odin-SMR). It is worth noting that the global-average concentration of N$_2$O estimated by each of the above mentioned instruments between 20~km and 40~km indicates that the observational uncertainty in the global amount of N$_2$O is about 10~\% \citep{Plieninger2016/ACP}. 

Lightning is not considered an important source of atmospheric N$_2$O, as shown in Table~11 of \cite{SchumannHuntrieser2007/SCP} where results from different studies and campaigns conclude that the global lightning-produced emission rate of N$_2$O is below 5~$\times$10$^{-4}$ ~Tg~N$_2$O-N~yr$^{-1}$.

Let us now turn to the global budget of NO and NO$_2$, which together make up NO$_x$. \cite{SchumannHuntrieser2007/SCP} presented an extensive study about the global production of NO$_x$ by lightning (or LNO$_x$) based on satellite and aircraft measurements, laboratory experiments and theoretical studies. 
Lightning is considered one of the major natural sources of atmospheric NO$_x$ emissions. Different studies estimate the global production of NO$_x$ in thunderstorms in a wide range between 1 and 20~Tg~NO-N~yr$^{-1}$  \citep{SchumannHuntrieser2007/SCP, huntrieser2016injection}. However, the most likely range is 5$\pm$3~Tg~N~yr$^{-1}$. Lightning would then contribute up to $\sim$10-15$\%$ of the total global emissions of  NO$_x$. It is probably something greater than 10\% now that anthropogenic emissions have decreased substantially in North America and Europe.

The uncertainties in the contribution of lightning to the global concentration of NO$_x$ is based on theoretical and empirical challenges. Laboratory results of NO$_x$ produced by electrical discharges are difficult to extrapolate to real lightning discharges, as both Cloud-to-Ground (CG) and Intra-Cloud (IC) lightning discharges are different from each other \citep{Price1997/JGR} and cannot be accurately reproduced in the laboratory. Space-based instruments measure NO$_2$ and cannot accurately measure the concentration of tropospheric NO$_x$ \citep{SchumannHuntrieser2007/SCP, beirle2010direct, bucsela2010lightning, pickering2016estimates}. This concentration has to be usually deduced from the concentration of other species that can react with NO$_x$ molecules. However, some uncertainties in the atmospheric chemical kinetics of NO$_x$ lead to imprecisions in the estimation of NO$_x$ from measurements. These estimations are often based on an assumed upper tropospheric (UT) chemical lifetime of NO$_x$ in a range between 2 and 8~days. 
Based on reanalysis of the measurements taken by the Deep Convective Clouds and Chemistry (DC3) atmospheric experiment, \cite{Nault2017/JGR} recently revised the interaction of atmospheric CH$_3$O$_2$NO$_2$ and HNO$_3$ with NO$_x$ molecules, estimating a new UT NO$_x$ lifetime of about 3 hours in the first few hours downwind of a thunderstorm instead of the previous scale of days.  Using this new analysis, \cite{Nault2017/JGR} estimated a global lightning production of NO$_x$ of about 9~Tg~NO-N~yr$^{-1}$. \cite{Nault2017/JGR} results indicate higher LNO$_x$ in the mid-latides than in the tropical regions, in agreement with \cite{SchumannHuntrieser2007/SCP}. However, the latest estimations of the global lightning NO$_x$ emissions by new cloud-sliced observations of UT NO$_2$ in the 6~km - 9~km range from the Ozone Monitoring Instrument (OMI) of the Aura mission combined with the GEOS-Chem model point to a global lightning NO$_x$ source of 5.5~Tg~N yr$^{-1}$ \citep{maraisnitrogen}. \cite{maraisnitrogen} reports no significant difference in LNO$_x$ production per flash between the tropics and mid-latitudes.
Stratospheric NO can cause ozone depletion through the processes \citep{crutzen1979/ARPS} 

\begin{linenomath*}
\begin{equation}
NO + O_3 \rightarrow NO_2 + O_2,
\end{equation}
\begin{equation}
NO_2 + O \rightarrow NO + O_2.
\end{equation}
\label{ozonedep}
\end{linenomath*}

Moreover, the oxidation of N$_2$O is the major source of stratospheric NO, producing 1~Tg~N~yr$^{-1}$ of NO$_x$ \citep{crutzen1979/ARPS}. Thus, the introduction of Blue Jets in global models as a new possible atmospheric source of N$_2$O and NO$_x$ could have a non negligible effect in the global budget of ozone.



\section{Model} 
\label{sec:models}

\subsection{WACCM4}
\label{sec:models}

The Whole Atmosphere Community Climate Model version 4 (WACCM4) \citep{Marsh2013/JC} is a global circulation model included in the Community Earth Climate System Model version 1 (CESM1). WACCM4 is an extension of the Community Atmosphere Model (CAM4) \citep{Marsh2013/JC, tilmes2015description, tilmes2016representation}. CAM4 couples the troposphere and the stratosphere chemistry, while WACCM4 extends up to the thermosphere. The chemistry of WACCM4 is based on version 4 of the Model for OZone And Related chemical Tracers (MOZART4) \citep{kinnison2007sensitivity, emmons2010description, lamarque2012/GMD, tilmes2015description}, including 183~species and 472~chemical reactions including gas-phase chemistry of neutrals and ions, photolysis and heterogenous chemistry.

We set the model domain extending from the surface to 140~km of altitude (5.96$\times$10$^{-6}$~hPa). We divide the vertical domain in 88 levels and set a horizontal resolution of 1.9$^{\circ}$ in latitude and 2.5$^{\circ}$ in longitude. We start the numerical experiment with WACCM4 running a complete year (from January 1999 to January 2000) without Blue Jets allowing free dynamics for each considered lightning parameterization. Then, we run the same period of time in the specified dynamics mode (SD-WACCM4) \citep{lamarque2012/GMD, smith2017/JAS}. In this study, we use the facility of SD but, instead of nudging to reanalysis fields, we nudge to the meteorological fields from a previous (free-running) WACCM simulation. The reason for using SD is to ensure that the basic dynamics in the lower and middle atmosphere is identical in simulations in which other changes are made. In this second run, temperature fields and horizontal winds in the troposphere and stratosphere are nudged at each model time step using the output of the first run. We apply the nudging from ground level to 80~km. The nudging is then tapered off in the ranges of altitude between 80 and 90~km, and finally removed at 90~km of altitude \citep{smith2017/JAS}.

Afterward, we use the same specified dynamics in order to run a complete year including all the combinations of lightning and Blue Jet parameterizations. This approach allows us to compare the simulations with and without Blue Jets in order to estimate their global chemical impact in the atmosphere.

As the lifetime of N$_2$O in the atmosphere is of the order of a century \citep{Prather2015/JGR}, we select the most realistic cases and repeat the process for a period of one decade. Following this approach, the obtained results would be closer to the chemical equilibrium. We discuss these cases in section~\ref{sec:results}.


\subsection{Lightning parameterizations}
\label{sec:lightning}

The temporal and geographical occurrence of Blue Jets obtained with WACCM4 will strongly depend on the global occurrence of lightning. In this section, we briefly highlight the particularities of each considered lightning parameterization.

The characteristic size of lightning is some orders of magnitude smaller than the WACCM4 grid size. Therefore, lightning are considered as sub-grid events in the model. WACCM4 includes a lightning parameterization based on the cloud top heights (CTH) \citep{Price1992/JGR} that estimates the density of lightning (or flashes) in each domain cell for every time step of 30~minutes. The regional and seasonal flash frequency produced by this parameterization roughly agrees with the observations recorded by the Lightning Imaging Sensor (LIS) and the Optical Transient Detector (OTD) \citep{Christian2003/JGR, cecil2014gridded} over a period of two decades. However, the implementation of this lightning parameterization in WACCM4 overestimates the total flashes per second taking place in the globe over a period of one year. According to OTD/LIS, the global lightning occurrence over a year is around 44~flashes per second, while this parameterization produces around 65~flashes per second. In addition, the lightning parameterization by \cite{Price1992/JGR} implemented in WACCM4 also underestimates the lightning occurrence over the oceans. For these reasons, the use of a parameterization for Blue Jets together with the CTH lightning parameterization by \cite{Price1992/JGR}  would probably underestimate the occurrence of Blue Jets over the oceans and would overestimate the global occurrence of Blue Jets. However, the spatial correlation between the flash frequency reported by OTD/LIS and the flash frequency estimated by CTH is 0.7602. This is the highest spatial correlation obtained by the use of different lightning parameterizations. Therefore, we choose the CTH lightning parameterization to show the primary results.

\cite{Allenp2002/JGR} developed a lightning parameterization based on the amount of convective precipitation (CP) over USA. We have tested this lightning parameterization in WACCM4, obtaining a good agreement between the predicted flash frequency (51~flashes per second) and the lightning occurrence reported by OTD/LIS. We obtain a spatial correlation between the flash frequency derived by CP and OTD/LIS of 0.5760. However, the implementation of this parameterization in WACCM4 produces a lightning occurrence that remains almost constant over the four seasons, in disagreement with observations. \cite{Allenp2002/JGR} also derived a lightning parameterization based on the upward mass flux (MFLUX) at 440~hPa. This parameterization produces again a good agreement between the predicted flash frequency (43~flashes per second) and the lightning occurrence reported by OTD/LIS. However, it slightly overestimates the occurrence of lightning in the oceans and in South America, while underestimates the flash density in some regions of Africa. In addition, MFLUX produces a low spatial correlation (0.4963) with the flash frequency reported by OTD/LIS.

Apart from these three ``classical" lightning parameterizations by \cite{Price1992/JGR}  and \cite{Allenp2002/JGR}, we also implement Blue Jet parameterizations in WACCM4 together with two, more recent, lightning parameterizations developed by \cite{Romps2014/SCI} (CPCAPE) and \cite{Finney2014/ACP} (ICEFLUX), respectively. The parameterization of  \cite{Romps2014/SCI}  is based on the precipitation rate and on the Convective Available Potential Energy (CAPE), while the parameterization by \cite{Finney2014/ACP} is based on the upward cloud ice flux at 440~hPa. 
The parameterization by \cite{Romps2014/SCI} produces global (52~flashes per second), regional and seasonal lightning frequencies that agree with the observation by OTD/LIS but it overestimates the flash occurrence over the oceans. The spatial correlation between the flash frequency derived by CPCAPE and the observations of OTD/LIS is 0.4540. The implementation of the parameterization developed by  \cite{Finney2014/ACP} underestimates by a factor of~2 the global lightning occurrence rate and results in a spatial correlation with the flash frequency reported by OTD/LIS of 0.6739.

\subsection{Blue Jet parameterizations in WACCM4}
\label{sec:bj}

\subsubsection{Blue Jet frequency}
\label{subsec:BJfrequency}
The characteristic size of Blue Jets is some orders of magnitude smaller than the horizontal size of WACCM4 grids. As in the case of lightning, Blue Jets have to be treated as sub-grid phenomena in WACCM4. Following the basic idea of the previously described global lightning parameterizations, we have developed two different Blue Jet parameterizations to be considered in global models. The first developed parameterization prescribes the estimation of global Blue Jets per minute to predict their chemical influence in the atmosphere. The second proposed parameterization is based on physical assumptions and does not impose the rate of occurrence of Blue Jets.

\subsubsection*{Parameterization based on ISUAL and the altitude of the tropopause (IS-TROP LOW / IS-TROP UP)}

\cite{Ignaccolo2006/GeoRL} proposed a formula to estimate the global rate of sprites based on reports of sprite detections. \cite{Ignaccolo2006/GeoRL} obtained a global occurrence rate of sprites about 2.8 per minute. According to ISUAL, the global occurrence rate of TLEs is around 4.13~per minute \cite{Chen2008/JGRA}, among which 3.23 are elves, 0.50 are sprites, 0.39 are halos and 0.01 are gigantic jets. As optical emissions from Blue Jets and lightning are difficult to separate, the global global occurrence rate of Blue Jets was not derived from ISUAL data. However, we assume that Blue Jets are less frequent than sprites and more frequent than Gigantic Jets. Therefore, as a first approximation we assume that the global occurrence rate of Blue Jet is in the range between 0.01 and 1.0 events per minute. Given that Blue Jets are triggered as a consequence of the remaining imbalance of charge in thunderclouds after lightning occurs \citep{Krehbiel2008/NatGe}, Blue Jet parameterizations must be spatially and temporally connected with any considered parameterization of lightning. Following these considerations, the total occurrence of Blue Jets at a given time would be the total occurrence of lightning flashes given multiplied by 3.6 $\times$ 10$^{-4}$ (UP) or 3.6 $\times$ 10$^{-6}$ (LOW), respectively. 

As we discussed in section~\ref{sec:bjstatistics}, the model proposed by \cite{Krehbiel2008/NatGe}  indicates that the inception of Blue Jets is favored when the two oppositely charged layers located in the upper part of thunderclouds do not mix. Hence, it is reasonable to expect the inception of Blue Jets when the convection near the cloud top is weak. In this regard, the Blue Jet reported by \cite{Chanrion2017/GRL} was triggered in a thundercloud whose top height was near the tropopause, where the lack of convection keeps the temperature relatively constant. WACCM4 and the most of Global Circulation Models can calculate the altitude of the tropopause and the cloud top height, two atmospheric variables that can be related with the possibility of Blue Jet inception \citep{Krehbiel2008/NatGe, Chanrion2017/GRL}. We can then distribute the predicted Blue Jets exclusively in the domain cells where the cloud top height is above the beginning of the tropopause or below tropopause by no more than one kilometer and there is lightning. We also impose as a condition to the existence of Blue Jets that the flash frequency is greater than zero in that domain cell. We restrict the locations where Blue Jets can be distributed to the range of latitudes between 60$^{\circ}$ S and 60$^{\circ}$ N. We name this Blue Jet parameterization as ``IS-TROP LOW" and ``IS-TROP UP", depending on whether the global occurrence rate of BJ is set to 3.6 $\times$ 10$^{-6}$ (LOW) or 3.6 $\times$ 10$^{-4}$ (UP) Blue Jets per lightning and where IS and TROP refer to ISUAL and to the height of the tropopause, respectively.  Although this parameterization could produce a realistic geographical occurrence of Blue Jets, the global occurrence rate is somehow imposed.



\subsubsection*{Parameterization based on lightning peak currents and the altitude of the tropopause (LPC-TROP LOW / LPC-TROP UP)}

According to the model developed by \cite{Krehbiel2008/NatGe}, a strong lightning discharge or a set of small amplitude CG lightning discharges occurring within a short time distance would probably precede the inception of a Blue Jet, since Blue Jets are produced by a large imbalance of charge. \cite{Chanrion2017/GRL} reported the observation of a Blue Jet preceded by a strong lightning discharge. Let us now use this observation to derive a more realistic Blue Jet parameterization based on the peak current value of the lightning possibly preceding a Blue Jet.
The Blue Jet reported by  \cite{Chanrion2017/GRL} was preceded by a negative Cloud-to-Ground (CG) lightning discharge with a peak current of  -167.5~kA. Elves, the less energetic TLEs, seem to be triggered by lightning discharges with peak currents whose absolute value is above 60~kA \citep{Barrington-Leigh1999/GeoRL/1}. As an approximation, we can assume that the peak current threshold of the lightning preceding Blue Jets is between 60~kA and 167.5~kA. We choose two representative values in this range to be the threshold of Blue Jets, such as 100~kA and 150~kA. According to the distribution of global lightning peak current reported by \cite{Said2013/JGR} using the Vaisala global lightning data set GLD360, approximately 1 \% of lightning have a peak current  above 100~kA and only 0.1 \% have a peak current  above 150~kA. We can then develop a Blue Jet parameterization in WACCM4 where the spatial occurrence of Blue Jets is again restricted to domain cells where the cloud top height is higher than one kilometer below the tropopause. However, instead of imposing the global occurrence of Blue Jets, we can now assume that the Blue Jet frequency in such domain cells is given by the amount of lightning in the domain cell with peak currents above 100~kA or 150~kA. Hence, we define the Blue Jet frequency in each cell where the cloud top height is near the tropopause as 0.01 (UP) or 0.001 (LOW) times the frequency of lightning. We refer to these two Blue Jet parameterizations as ``LPC-TROP UP" and ``LPC-TROP LOW", where LPC and TROP recall to lightning peak current and to the height of the tropopause, respectively. The maximum peak current of lightning is not homogeneously distributed over land and ocean \citep{Said2013/JGR}. However, we do not include in this parameterization any parameter to take into account this inhomogeneity. This simplification is justified because the scope of this paper is to describe the global chemical influence of Blue Jets rather that the regional influence.

\subsubsection{Chemical impact of Blue Jets}
\label{subsec:chemicalimpact}
\cite{Winkler2015/JASTP} developed the most detailed zero-dimensional model until now to predict the local chemical impact of a Blue Jet in the center of Blue Jet leader and streamers. \cite{Winkler2015/JASTP} estimated the chemical impact of an upward propagating Blue Jet at different altitudes between 18~km and 38~km, obtaining a significant enhancement in the densities of some chemical species such as N$_2$O, NO or O and a decrease in the density of O$_3$ in the center of Blue Jets. We use the local chemical impact of a single Blue Jet obtained by \cite{Winkler2015/JASTP}  together with the previously derived Blue Jet parameterizations to estimate the global chemical impact of Blue Jets using WACCM4. We assume that each Blue Jet would produce an enhancement in the concentrations of N$_2$O, NO or O.  However, as Blue Jets are considered as sub-grid phenomena in WACCM4, we take into account the following considerations:

\begin{enumerate}
\item As the area of a WACCM4 cell is larger than the horizontal cross-section of the Blue Jet, we have to estimate the total number of species produced by all Blue Jets at each altitude and distribute them over the area of the grid at each altitude level. Hence, we need to estimate the electrodynamical radius of the Blue Jet where the chemical reactions are produced. \cite{Winkler2015/JASTP} noted that while the optical radius of a Blue Jet is a few hundreds of meters \citep{Wescott2001/JGR}, the electrodynamical radius could be between 20 and 100~times smaller \citep{Shneider2012/PhysPlas, Milikh2014/JGR}. Based on optical observations by \cite{Wescott2001/JGR}, we have assumed that Blue Jets have an optical radius of 250~m at their base. Hence, the electrodynamical radius $R_e$ would be in the range between $R_1$ = 2.5~m and $R_2$ = 12.5~m. According to  \cite{Winkler2015/JASTP}, the production of N$_2$O and NO is dominated by the leader for altitudes ranging between 18~km and 28~km and by streamers between 28~km and 38~km. We have then assumed that the electrodynamical radius of the Blue Jet is completely filled by a leader between 18~km and 28~km of altitude and by streamers between 28~km and 38~km of altitude.

The global production of N$_2$O by Blue Jets can then be estimated using the density variations reported by \cite{Winkler2015/JASTP} (figure~19 of \cite{Winkler2015/JASTP}). As a first approximation, we can consider a Blue Jet as a 20~km long cylinder with a constant radii of 2.5~m or 12.5~m formed by a leader and a streamer region. A single Blue Jet would produce between 2 $\times$ 10$^{28}$ and 6 $\times$ 10$^{29}$ molecules of N$_2$O for electrodynamical radii of 2.5~m and 12.5~m, respectively. Assuming that the global occurrence rate of Blue Jets is between 0.01 and 1~per minute \citep{Ignaccolo2006/GeoRL, Chen2008/JGRA}, we find that Blue Jets with a radii of 2.5~m would produce between 6 $\times$ 10$^{-3}$~Tg~N$_2$O-N~yr$^{-1}$ and 0.6~Tg~N$_2$O-N~yr$^{-1}$, while Blue Jets with a radii of 12.5~m would produce between 0.15~Tg~N$_2$O-N~yr$^{-1}$ and 15 ~Tg~N$_2$O-N~yr$^{-1}$.

\item We assume that the production of species decays parabolically across the radial coordinate $r$ from the center of the Blue Jet up to the limit of the electrodynamical radius as 

\begin{linenomath*}
\begin{equation}
N(r) = N_{max} \left(1 - \frac{r^2}{R^2_e} \right),
\label{NR}
\end{equation}
\end{linenomath*}

where $R_e$ is the electrodynamical radius and $N_{max}$ is the enhancement in the density of species in the symmetry axis of the Blue Jet as predicted by \cite{Winkler2015/JASTP}.


\item Recorded optical emissions from Blue Jets indicate an increase of its radius with altitude \citep{Wescott2001/JGR}. We assume that the Blue Jet radius increases with altitude following the simple scale law

\begin{linenomath*}
\begin{equation}
P_i R_i = P_j R_j,
\label{scalelay}
\end{equation}
\end{linenomath*}

where $P$ and $R$ are the atmospheric pressure and Blue Jet radius at two different altitude levels denoted as $i$ and $j$.


\end{enumerate}

Following these considerations, the Blue Jet region at 30~km of altitude filled by streamers would have a radius between 16~m and 80~m. Given the assumed altitude-dependence of the leader and streamer-region radius and the production profile by \cite{Winkler2015/JASTP}, the total production of N$_2$O and NO by a Blue Jet is dominated by the leader phase.


\section{Results} 
\label{sec:results}

We have implemented in WACCM4 the Blue Jet parameterizations derived in section~\ref{sec:bj} using five different lightning parameterizations. We present the obtained global occurrence rate of Blue Jets in subsection~\ref{sec:occurrence}. We have coupled these Blue Jet frequencies with the chemical impact of a single Blue Jet predicted by \cite{Winkler2015/JASTP}. In subsection~\ref{sec:gchemical} we present the predicted global impact of Blue Jets for each global parameterization.

\subsection{Global occurrence and seasonal cycle of Blue Jets}
\label{sec:occurrence}

\begin{figure}
\includegraphics[width=1.0\columnwidth]{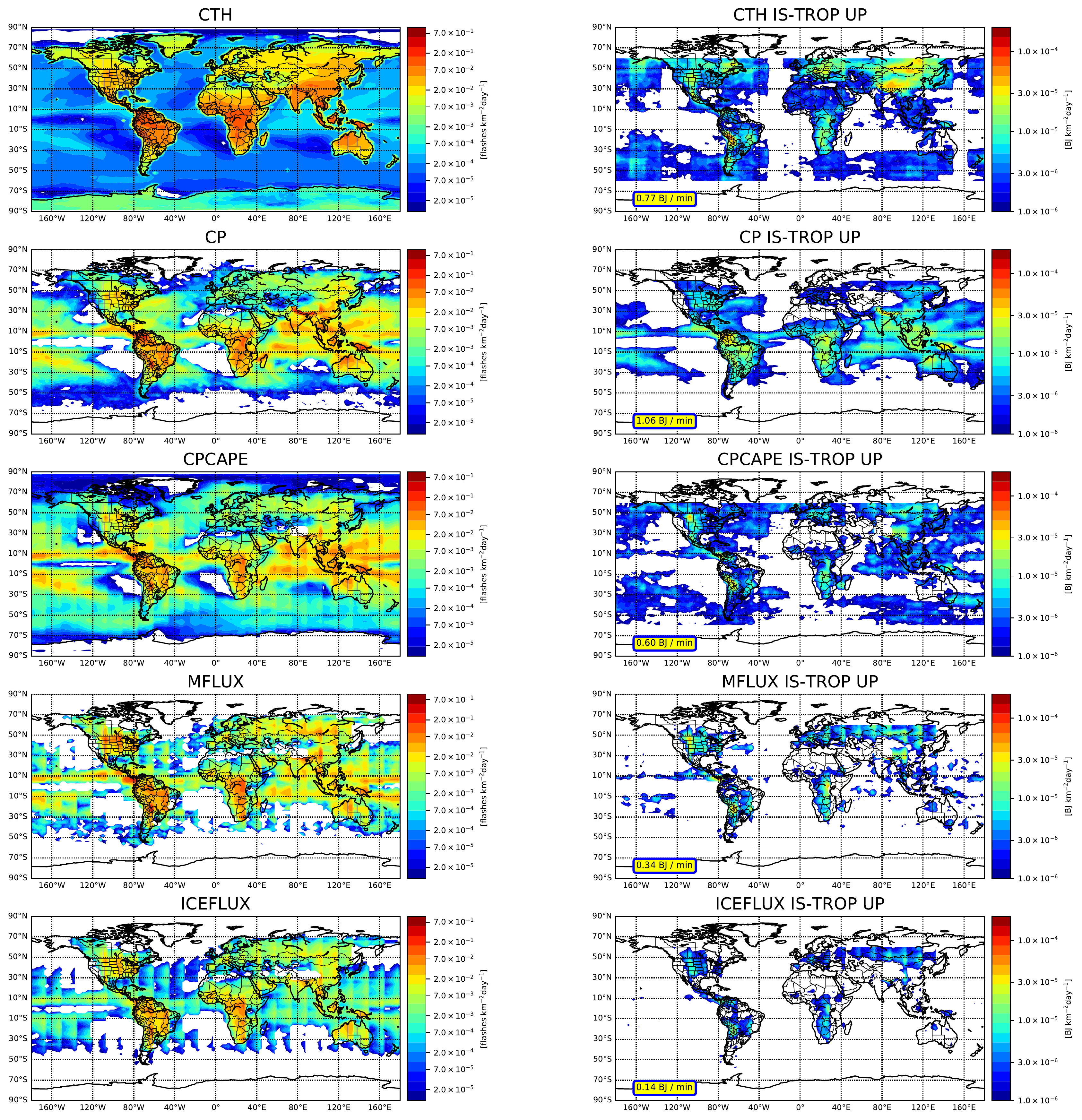}
\caption{\label{fig:bj_cases_1_2}
(First column) annual average lightning flash frequencies in flashes km$^{-2}$day$^{-1}$ and (second column) annual average Blue Jet frequencies in BJ km$^{-2}$day$^{-1}$ using the Blue Jet parameterization denoted as ``IS-TROP LOW" and different lightning parameterizations. We have used different lightning parameterizations denoted as CTH \citep{Price1992/JGR} based on the cloud top height, CP \citep{Allenp2002/JGR} based on the precipitation rate, CPCAPE \citep{Romps2014/SCI} based on the precipitation rate and convective available potential energy (CAPE), MFLUX \cite{Allenp2002/JGR} based on the updraft mass flux and ICEFLUX \citep{Finney2014/ACP} based on the upward cloud ice flux. We annotate in boxes the total annual Blue Jets per minute.
}
\end{figure}

\begin{figure}
\includegraphics[width=1.0\columnwidth]{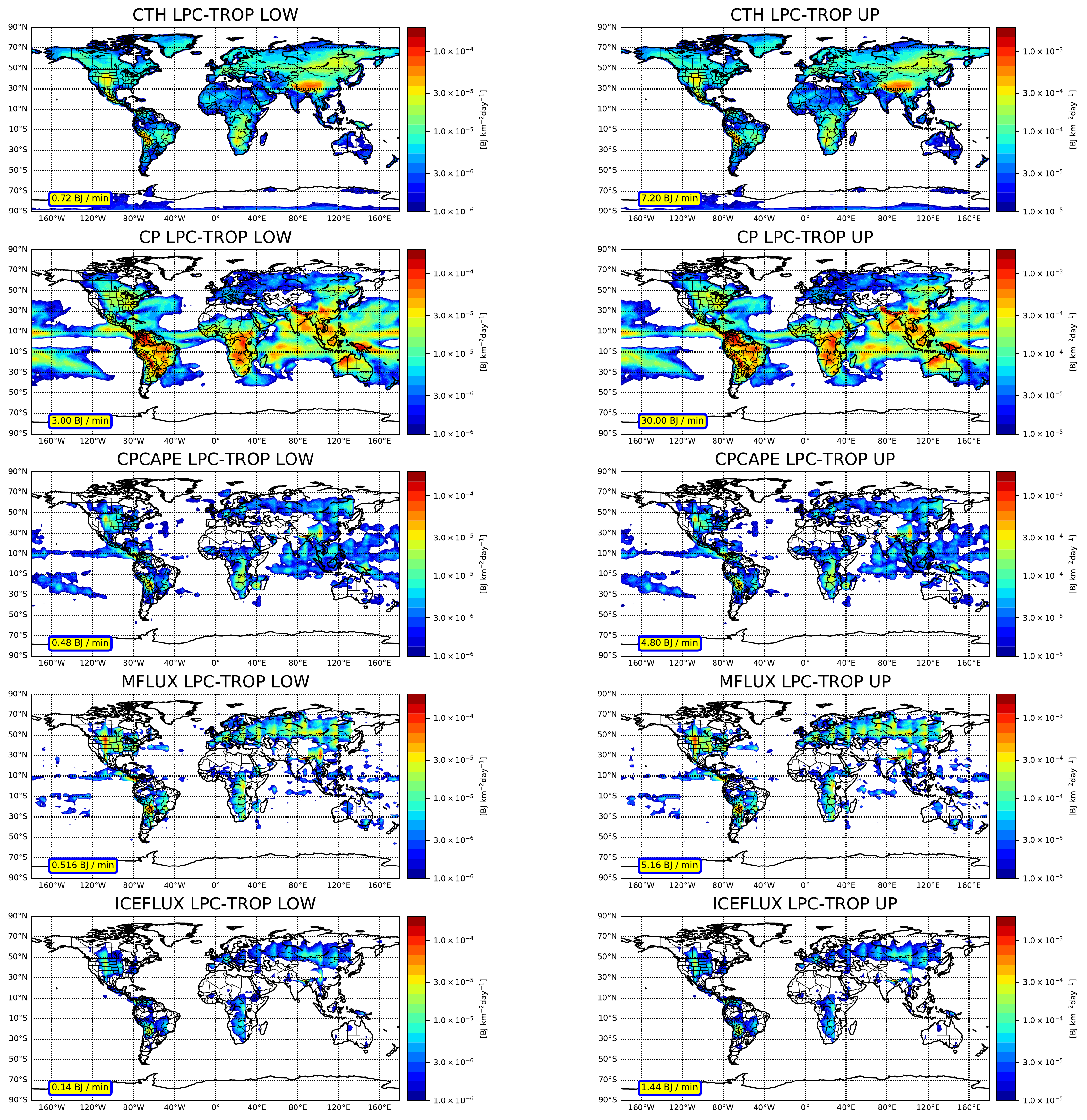}
\footnotesize
\caption{\label{fig:bj_cases_3}
Annual average Blue Jet frequencies BJ km$^{-2}$day$^{-1}$ using the parameterizations denoted as ``LPC-TROP LOW" and ``LPC-TROP UP". The shown Blue Jet frequencies have been calculated using different lightning parameterizations denoted as CTH \citep{Price1992/JGR} based on the cloud top height , CP \citep{Allenp2002/JGR} based on the precipitation rate, CPCAPE \citep{Romps2014/SCI} based on the precipitation rate and convective available potential energy (CAPE), MFLUX \citep{Allenp2002/JGR} based on the updraft mass flux and ICEFLUX \citep{Finney2014/ACP} based on the upward cloud ice flux. We annotate in boxes the total annual Blue Jets per minute.
\normalsize
}
\end{figure}

Let us firstly focus on the global occurrence of Blue Jets derived for each combination of lightning and Blue Jet parameterizations. The first column of figure~\ref{fig:bj_cases_1_2} shows the obtained lightning flash frequency using different lightning parameterizations in WACCM4. The second column of figure~\ref{fig:bj_cases_1_2} shows the Blue Jet frequency for the Blue Jet parameterization denoted as ``IS-TROP LOW" using different lightning parameterizations. As we have previously detailed, this Blue Jet parameterization take as input the global occurrence rate of Blue Jet estimated from the TLE occurrence reported by ISUAL. The Blue Jet frequency obtained with the BJ parameterizations ``LPC-TROP LOW" and ``LPC-TROP UP" using different lightning parameterizations is plotted in figure~\ref{fig:bj_cases_3}. In the ``LPC-TROP" parameterization, the Blue Jet occurrence rate is not fixed to any given value. Instead, it is based on the physical assumptions previously described in section~\ref{sec:bj}.

The Blue Jet frequencies obtained with all the considered Blue Jet parameterizations and with the CTH lightning parameterization are collected in table~\ref{tab:results}. The results obtained with the other tested lightning parameterizations are shown in the supplementary material.

The Blue Jet parameretizations ``IS-TROP" and the most of the Blue Jet parameterizations ``LPC-TROP LOW" produce a global occurrence rate of Blue Jets lower than 1 BJ per minute, as estimated from the TLE frequency reported by ISUAL. However, the Blue Jet parameterizations ``LPC-TROP UP" and ``CP LPC-TROP LOW" significantly overestimate the Blue Jet frequency. 

The comparison of the spatial distribution of lightning flashes and Blue Jets indicates that the relative occurrence of Blue Jets in Asia with respect to other regions is larger than the relative occurrence of lightning flashes. In addition, most of the considered parameterizations produce a maximum in the lightning flash frequency and in the Blue Jet frequency over Africa and in the north of South America. 

\begin{figure}
\includegraphics[width=0.8\columnwidth]{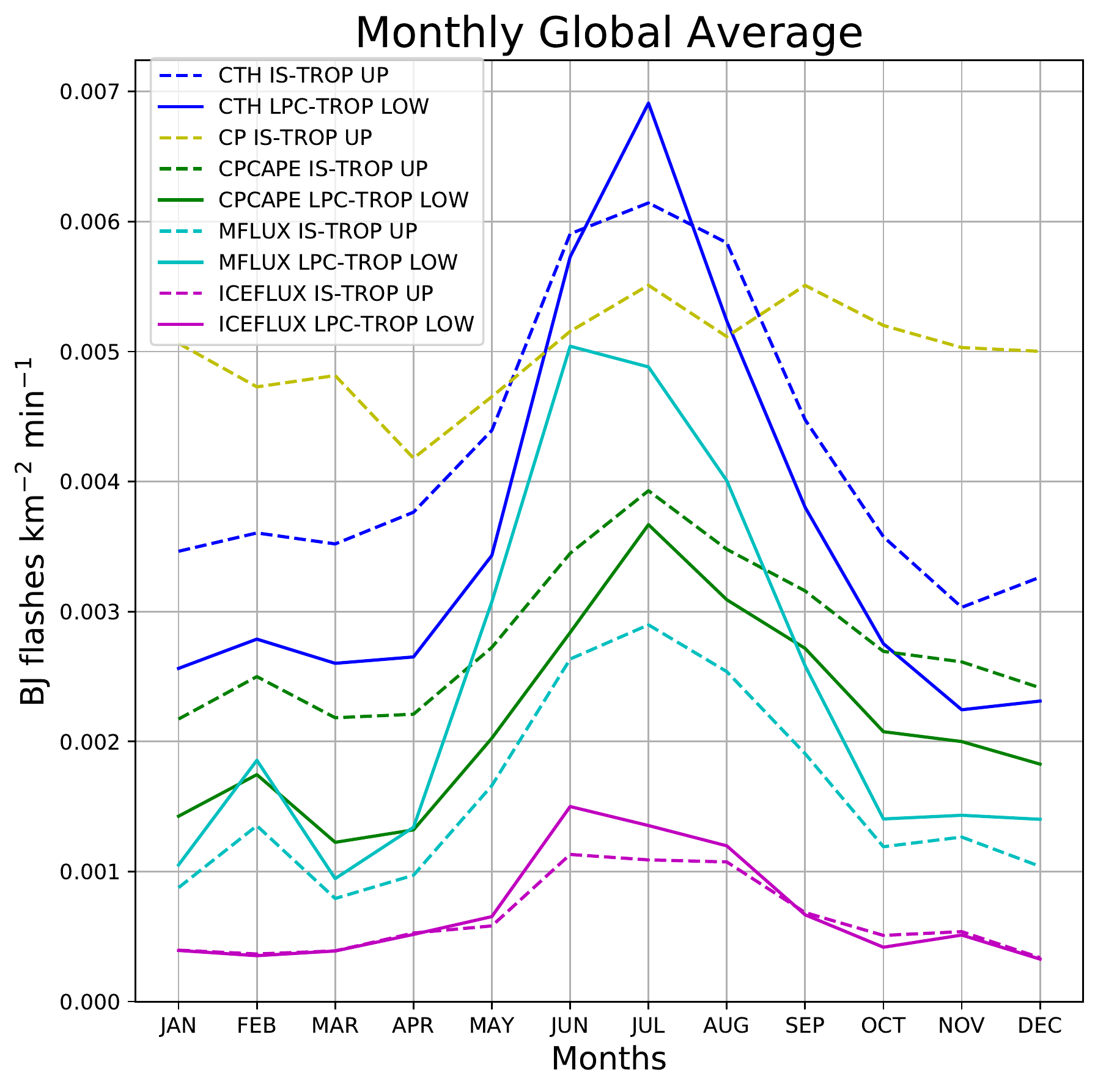}
\caption{\label{fig:bj_monthly}
Monthly global average Blue Jet frequencies BJ km$^{-2}$day$^{-1}$ using the developed Blue Jets parameterization. The shown Blue Jet frequencies have been calculated using different lightning parameterizations denoted as CTH \citep{Price1992/JGR} based on the cloud top height, CP \citep{Allenp2002/JGR} based on the precipitation rate, CPCAPE \citep{Romps2014/SCI} based on the precipitation rate and convective available potential energy (CAPE), MFLUX \citep{Allenp2002/JGR} based on the updraft mass flux and ICEFLUX \citep{Finney2014/ACP} based on the upward cloud ice flux.
}
\end{figure}

The monthly global average occurrences of Blue Jets obtained with different lightning parameterization schemes are presented in figure~\ref{fig:bj_monthly}. The maximum occurrence of Blue Jets takes place between June and August, coinciding with the maximum occurrence of lightning. As the occurrence of Blue Jets is related with a high lightning activity on clouds \citep{Krehbiel2008/NatGe} (see figure~\ref{fig:BJ_wescott}), we conclude that the obtained coincidence of the seasonal cycle of lightning and Blue Jets can be considered as realistic.

\subsection{Global chemical impact of Blue Jets}
\label{sec:gchemical}

We follow the simulation scheme proposed in section~\ref{sec:models} in order to predict the global chemical impact of Blue Jets in the atmosphere. We use the same specified dynamics to simulate the atmosphere with and without Blue Jets. 

First, we run simulations of one year for all the considered Blue Jet parameterizations (subsection~\ref{sec:chem1}). Then, we choose some of the most representative cases and extend the simulations up to ten years in order to obtain the chemical influence of Blue Jets when equilibrium is reached  (subsection~\ref{sec:chem10}). We do not include in this discussion the chemical impact of Blue Jets using the parameterizations that overestimate the Blue Jet frequency (``LPC-TROP UP"). 

\subsubsection{Transient response}
\label{sec:chem1}
In this section we present and discuss the global impact of Blue Jets over one year for all the considered cases. It is important to note that simulations of one year including the chemical perturbation of Blue Jets do not allow the atmosphere to reach the equilibrium. However, these short simulations are useful to choose the most realistic cases before extending the simulation to five and ten years. 
We collect in table~\ref{tab:results} the total annual production, i.e., the total number of NO, N$_2$O and O molecules injected in the atmosphere by Blue Jets for all the considered cases of the CTH lightning parameterization, while we show in the supplementary material the results corresponding to other lightning parameterizations. The cases in which the production of N$_2$O is larger than the natural source of atmospheric N$_2$O (10.2~Tg~N$_2$O-N~yr$^{-1}$)  \citep{davidson2009/nat, Prather2015/JGR} and the total occurrence rate of Blue Jets is higher than 1 BJ per minute will be considered as unrealistic scenarios. Therefore, the realistic scenarios would be most of the``IS-TROP UP", all ``IS-TROP LOW" and most of ``LPC-TROP LOW". The lower realistic scenario (3.6 $\times$ 10$^{-6}$ BJ per lightning flash and R$_1$ = 2.5~m) produces a Blue Jet frequency of 1.4 $\times$ 10$^{-3}$ BJ per minute and 6.6 $\times$ 10$^{-4}$ Tg~N$_2$O-N~yr$^{-1}$ (ICEFLUX IS-TROP LOW R$_1$), while the higher realistic case (3.6 $\times$ 10$^{-4}$ BJ per lightning flash and R$_2$ = 12.5~m) produces a Blue Jet frequency of 0.72~BJ per minute and 7.6~Tg~N$_2$O-N~yr$^{-1}$ (CTH LPC-TROP LOW R$_2$). 
The predicted production of NO in the so-called realistic cases is about two orders of magnitude lower that the production of NO by lightning. The global production of NO by Blue Jets is then negligible. The global production of O is also negligible.

\begin{figure}
\includegraphics[width=1\columnwidth]{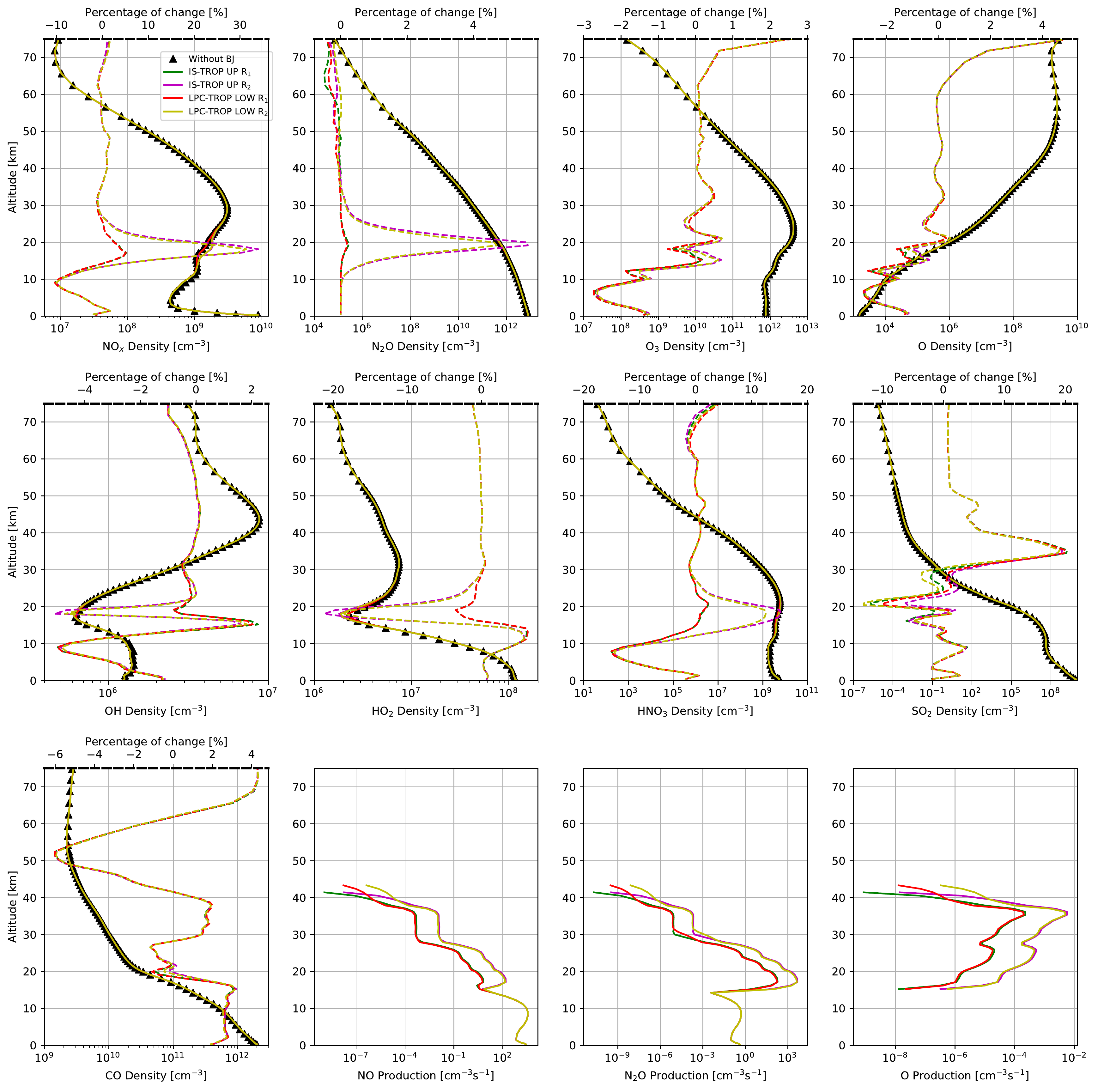}
\caption{\label{fig:bj_price}
Solid lines correspond to annual global average density of some species after a WACCM4 simulation of 1~year including Blue Jets and using the lightning parameterization CTH \citep{Price1992/JGR}. Triangles correspond to the same simulation with lightning but without Blue Jets. Dashed lines represent the percentage difference when Blue Jets are included. The last three subplots in the lower row show the total production rate of NO, N$_2$O and O, respectively by Blue Jets and lightning.
}
\end{figure}

Let us now estimate the transient chemical impact of Blue Jets in the atmosphere over one year. For this purpose, we calculate the global annual average vertical profile of some chemical species obtained from the simulations of Blue Jets and compare them with the profiles produced in the simulations without Blue Jets. We plot the obtained results with the cases whose predicted Blue Jet frequency is close to the maximum value estimated from the TLE occurrence reported by ISUAL (IS-TROP UP R$_1$ and R$_2$; and LPC-TROP LOW R$_1$ and R$_2$, respectively) in figure~\ref{fig:bj_price}, together with the percentage of change at each altitude between simulations with and without Blue Jets (relative enhancement). The last three plots of figure~\ref{fig:bj_price} correspond to the vertical production rate of NO, N$_2$O and O by both Blue Jets and lightning. Figure~\ref{fig:bj_price} shows results for the CTH lightning parameterization, while the supplementary material shows figures collecting results with other lightning parameterizations (CP, CPCAPE, MFLUX, ICEFLUX).


It is worth analyzing the obtained chemical impact for each considered species. The most remarkable chemical impact of Blue Jets are the enhancements in the densities of NO$_x$ and N$_2$O at altitudes between 10~km and 30~km. Most of the simulations producing a realistic Blue Jet frequency and imposing R$_2$ = 12.5~m (IS-TROP UP R$_2$ and LPC-TROP LOW R$_2$) predict maximum density increases of NO$_x$ and N$_2$O of 30~\% and 5~\%, respectively. Simulations that produce the lowest possible Blue Jet frequency (IS-TROP LOW R$_1$ and R$_2$) and simulations producing a realistic Blue Jet frequency (IS-TROP UP R$_1$ and LPC-TROP LOW R$_1$) and imposing R$_1$ = 2.5~m predict a negligible influence of Blue Jets in the global amount of NO$_x$ and N$_2$O.  

Vertical profiles of other species can also be influenced by the inclusion of Blue Jets in the global atmospheric chemistry. Blue Jets could produce a decrease in the upper tropospheric density of OH and HO$_2$  of about 5~\% and 20~\%, respectively. The injected NO molecules would led to an increase in OH and a reduction of HO$_2$ by the process NO + HO$_2$ $\rightarrow$ NO$_2$ + OH \citep{murray2013interannual}. However, the conversion of NO into NO$_2$ can also contribute to a decrease in the concentration of OH, specially at lower altitudes by the process NO$_2$ + OH + M $\rightarrow$ HNO$_3$ + M, where M represents air molecules (N$_2$ + O$_2$). According to our results, the concentration of HNO$_3$ and SO$_2$ could increase about 20~\%. HNO$_3$ and SO$_2$ molecules can directly contribute to the production of acid rain \citep{seinfeld2016atmospheric}. The density profile of CO can exhibit both relative increases and decreases at different altitudes, as its gains and losses mechanisms depend on the concentration of OH according to the process CO + OH $\rightarrow$ CO$_2$ + H \citep{murray2013interannual}.
The global density profiles of other species, such as O$_3$ and O, are not significantly influenced by Blue Jets.

\begin{figure}
\includegraphics[width=1\columnwidth]{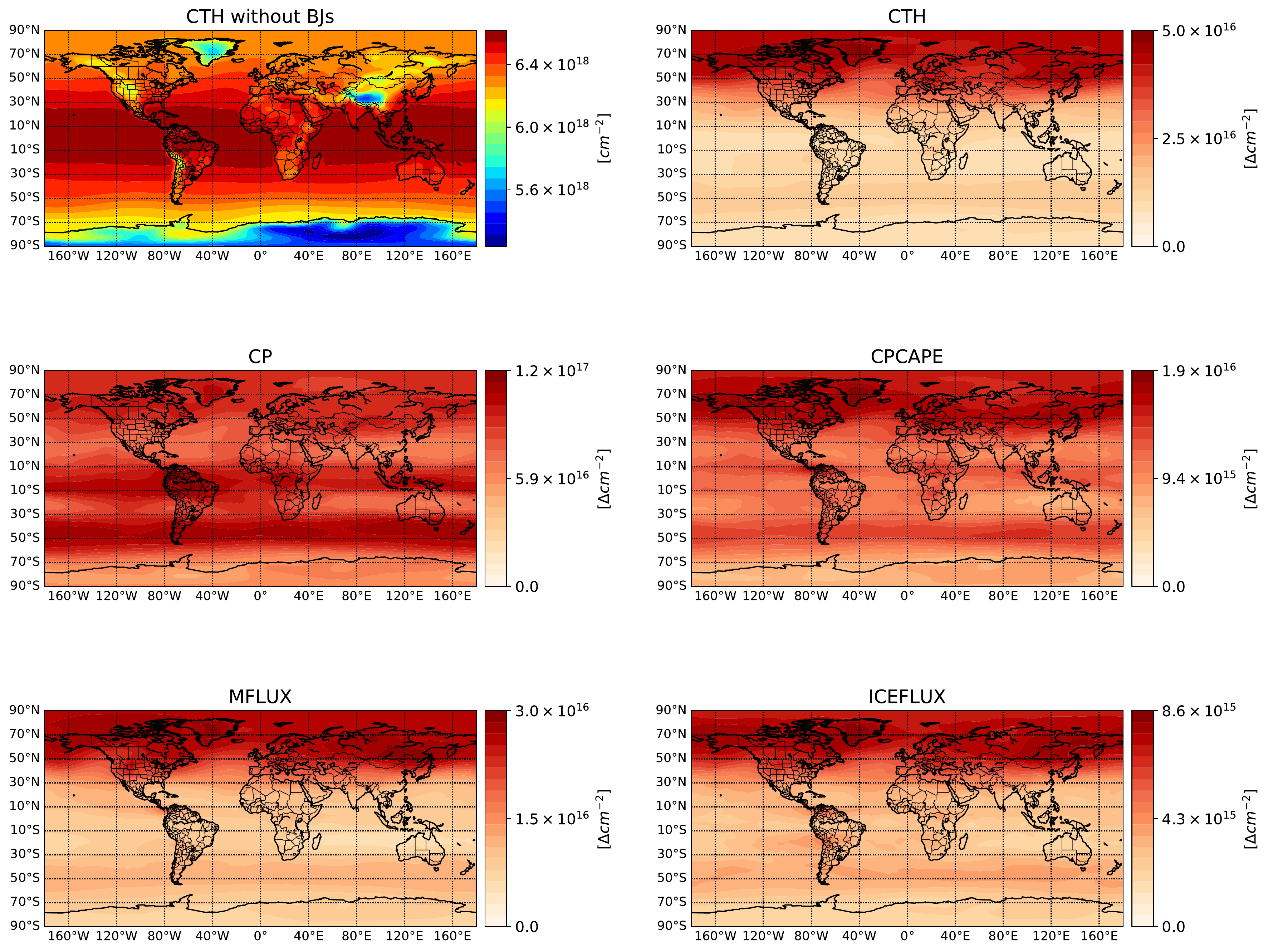}
\caption{\label{fig:bj_column_chem_enhancement_cases}
The top left panel shows the annual average total column density of N$_2$O after a WACCM4 simulation of 1~year using the lightning parameterization by \cite{Price1992/JGR} without Blue Jets. The other panels correspond to the variation in the annual average total column density of N$_2$O between two simulations of 1~year with and without Blue Jets using different lightning parameterizations and the realistic Blue Jet parameterization ``LPC-TROP UP R$_1$" . 
}
\end{figure}

The first panel of figure~\ref{fig:bj_column_chem_enhancement_cases} shows the annual average total column density of N$_2$O after a WACCM4 simulation of 1~year using the lightning parameterization by \cite{Price1992/JGR} without Blue Jets. The rest of the panels in figure~\ref{fig:bj_column_chem_enhancement_cases} show the annual average total column density difference of N$_2$O between two simulations of 1~year with and without Blue Jets using different lightning parameterizations. The geographical distribution of N$_2$O changes is directly linked to the adopted lightning parameterization (see figure~\ref{fig:bj_cases_3}), with strong increases in N$_2$O in the tropics and/or mid-latitudes in relation to a local stronger Blue Jet occurrence in those regions. Interestingly, all lightning parameterizations produce an enhancement in the concentration of N$_2$O near the northern high latitude and polar regions. Given the limited amount of Blue Jet simulated to occur at high latitude, N$_2$O is likely increased by wave-driven transport and mixing from lower latitudes in the extratropical upper troposphere-lowermost stratosphere on relatively fast timescales (see e.g. \cite{holton1995stratosphere}). On longer timescales (see e.g. the 5 and 10 years cases presented below), increases in N$_2$O at high latitude can occur through poleward and downward adiabatic transport of tropical air. Similar effects are produced also in simulations of LNO$_x$ by \cite{grewe2009impact} with high impact of LNO$_x$ to changes at high latitude.



\subsubsection{Response close to equilibrium}
\label{sec:chem10}

The analysis of the global chemical impact of Blue Jets presented in the previous section is based on a simulation of one year. As we pointed out in section~\ref{sec:models}, the lifetime of N$_2$O is of the order of a century, while the time scale of the overturning circulation is about 5~years.

\begin{figure}
\includegraphics[width=1\columnwidth]{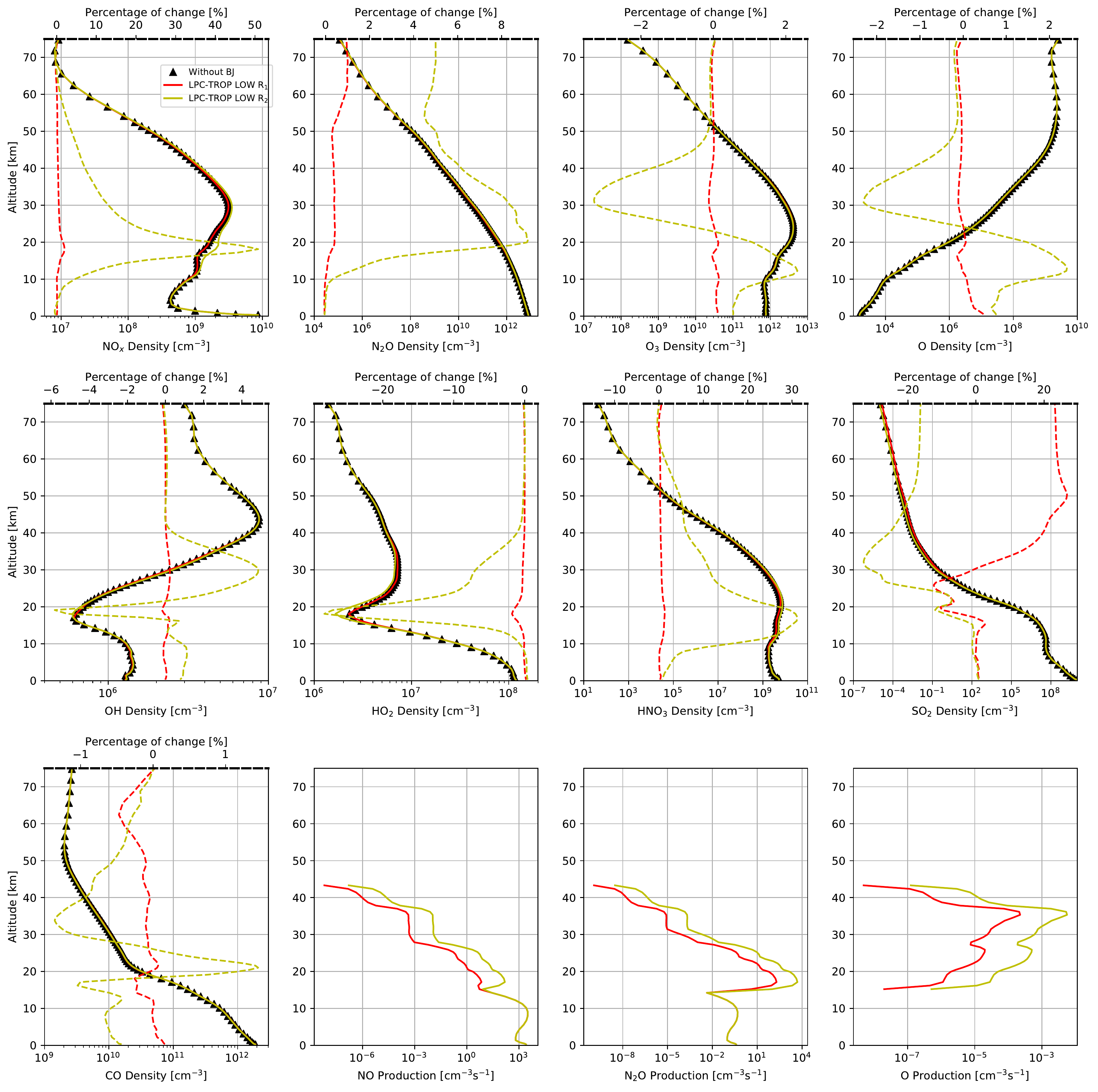}
\caption{\label{fig:bj_price_2004}
Solid lines correspond to annual global average density of some species after a WACCM4 simulation of 5~years including Blue Jets and using the lightning parameterization CTH \citep{Price1992/JGR}. Triangles correspond to the same simulation with lightning but without Blue Jets. Dashed lines represent the percentage variation with respect to a similar simulation without Blue Jets. The last three subplots in the lower row show the total production rate of NO, N$_2$O and O by lightning and Blue Jets. Note that the horizontal upper scale of figures~\ref{fig:bj_price}, \ref{fig:bj_price_2004} and \ref{fig:bj_price_2009} are different.
}
\end{figure}

\begin{figure}
\includegraphics[width=1\columnwidth]{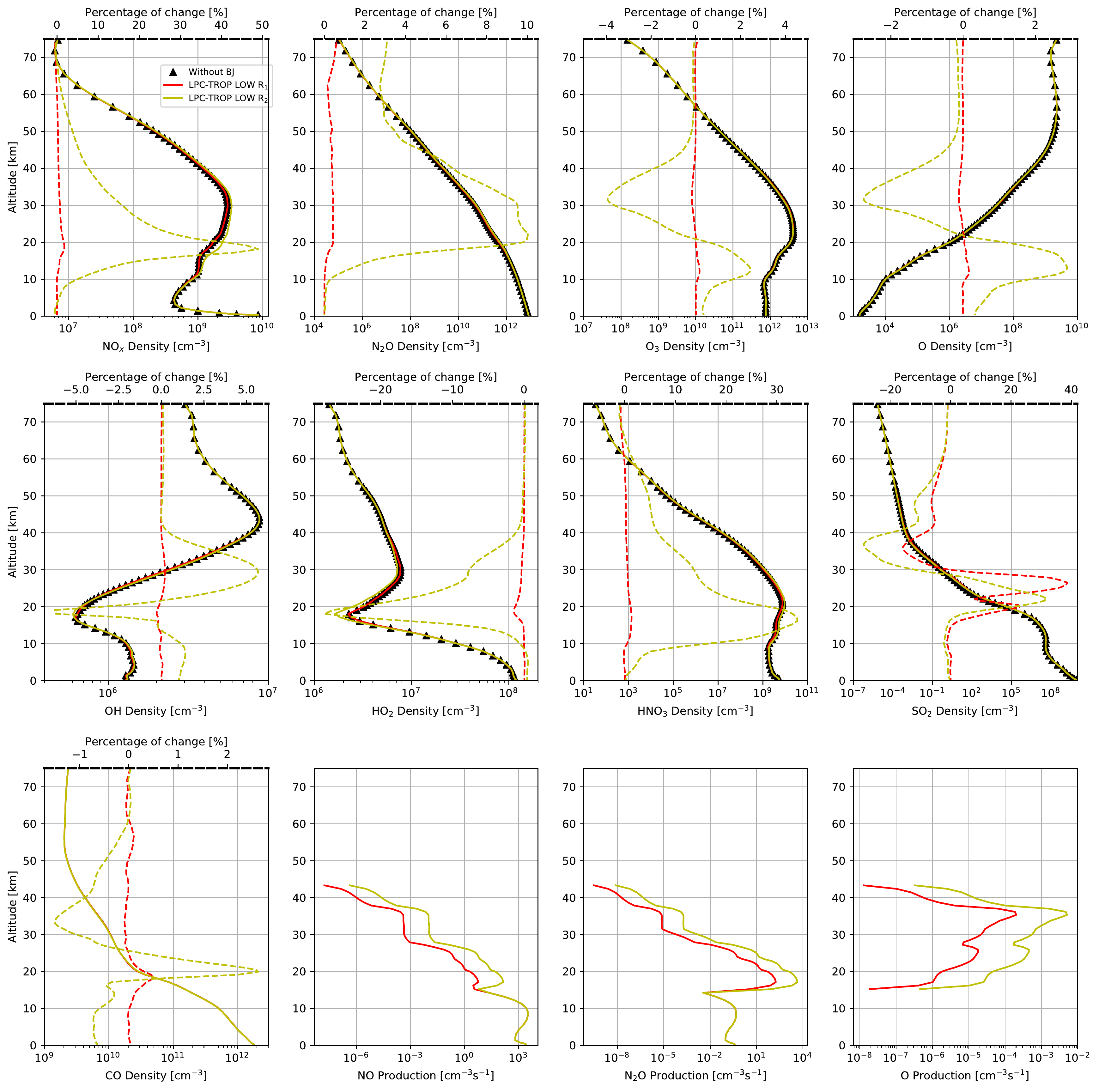}
\caption{\label{fig:bj_price_2009}
Solid lines correspond to annual global average density of some species after a WACCM4 simulation of 10~year including Blue Jets and using the lightning parameterization  CTH \citep{Price1992/JGR}. Triangles correspond to the same simulation with lightning but without Blue Jets. Dashed lines represent the percentage variation with respect to a similar simulation without Blue Jets. The last three subplots in the lower row show the total production rate of NO, N$_2$O and O by lightning and Blue Jets. Note that the horizontal upper scale of figures~\ref{fig:bj_price}, \ref{fig:bj_price_2004} and \ref{fig:bj_price_2009} are different.
}
\end{figure}

We select two of the previously identified realistic cases in terms of Blue Jet frequency (LPC-TROP LOW R$_1$ and R$_2$) and extend the CTH-based simulations up to five and ten years. We also extend the control case without Blue Jets up to five and ten years. This approach allows us to see the global distribution after the injected species have been transported. It is important to emphasize that a simulation of more than 100~years would be necessary to reach the complete chemical equilibrium. However, such a long simulation is out of the scope of this paper. We plot on figures~\ref{fig:bj_price_2004} and~\ref{fig:bj_price_2009} the atmospheric chemical influence of Blue Jets annual averaged for the fifth and the tenth year of simulation, respectively. The chemical influence of Blue Jets is a factor of two larger in the five year simulation (see figure~\ref{fig:bj_price_2004}) than in the simulation of one year (figure~\ref{fig:bj_price}). Hence, we cannot assume that the atmosphere has already reached an equilibrium after including Blue Jets. However, the chemical influence of Blue Jets as shown in the 10 year simulation (see figure~\ref{fig:bj_price_2009}) is quite similar to the one obtained in the simulation of five years, indicating that a simulation of ten years may be sufficient to estimate the global chemical impact of Blue Jets despite the 100~year lifetime of N$_2$O. After a simulation of ten years, the density enhancements and decreases obtained in the previous section (one year simulations) are increased by a factor of two. The increase in the tropospheric density of HNO$_3$ suggests that Blue Jets could also have a direct influence in the acidity of rainwater.

\begin{figure}
\includegraphics[width=1\columnwidth]{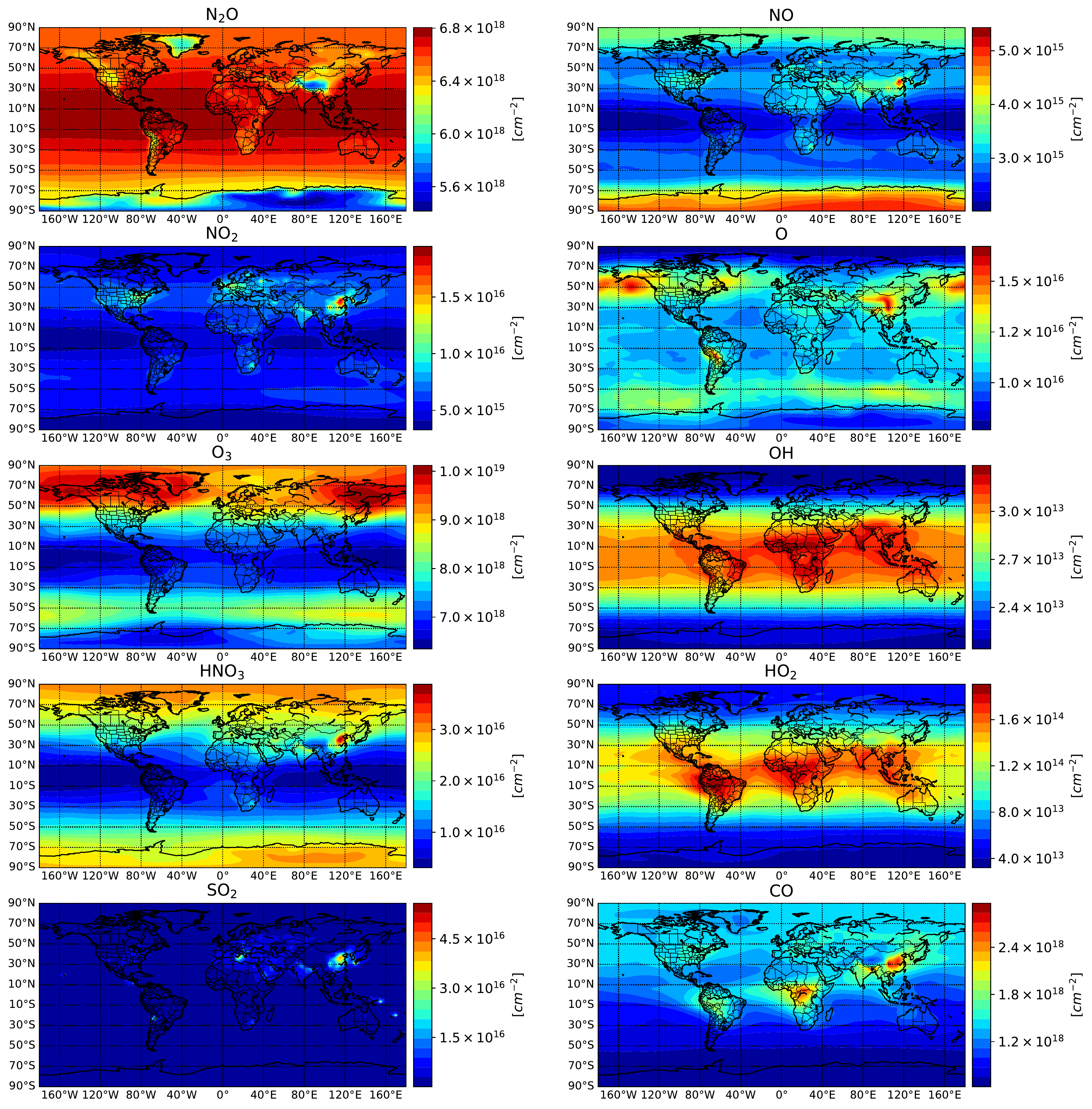}
\caption{\label{fig:bj_column_chem_absolute}
Annual average total column density of some chemical species after a WACCM4 simulation of 10~years including Blue Jets. These subplots have been calculated using the lightning parameterization based on the cloud-top height CTH \citep{Price1992/JGR} and the Blue Jets parameterization denoted as ``LPC-TROP LOW R$_2$".
}
\end{figure}

\begin{figure}
\includegraphics[width=1\columnwidth]{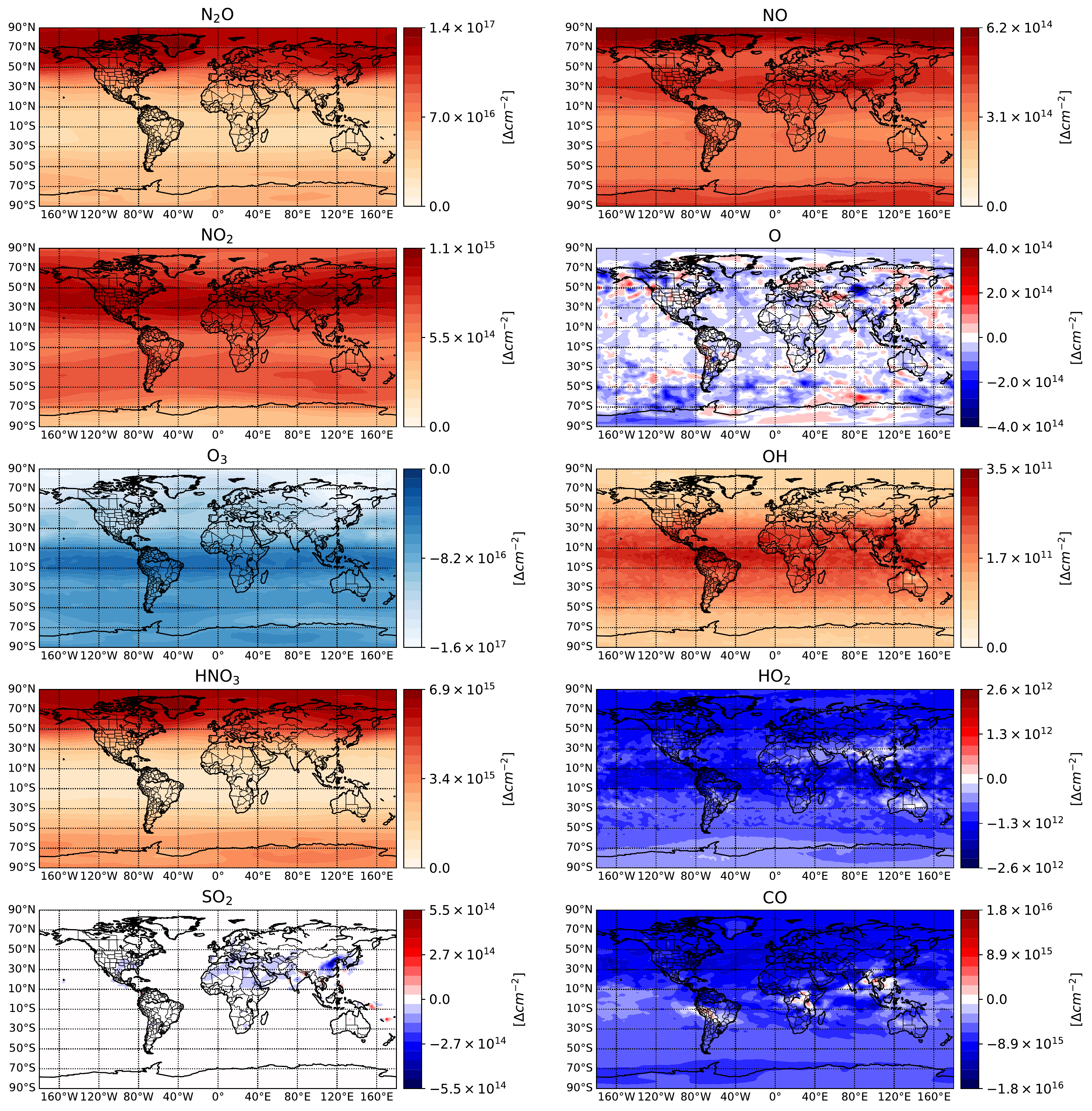}
\caption{\label{fig:bj_column_chem_enhancement}
Differences in the annual average total column density of some chemical species between two simulations of 10~years with (as in figure~\ref{fig:bj_column_chem_absolute}) and without Blue Jets. Positive values correspond to enhancement in densities due to Blue Jets, while negative variations represent density decrease produced by Blue Jets. 
}
\end{figure}

\begin{figure}
\includegraphics[width=1\columnwidth]{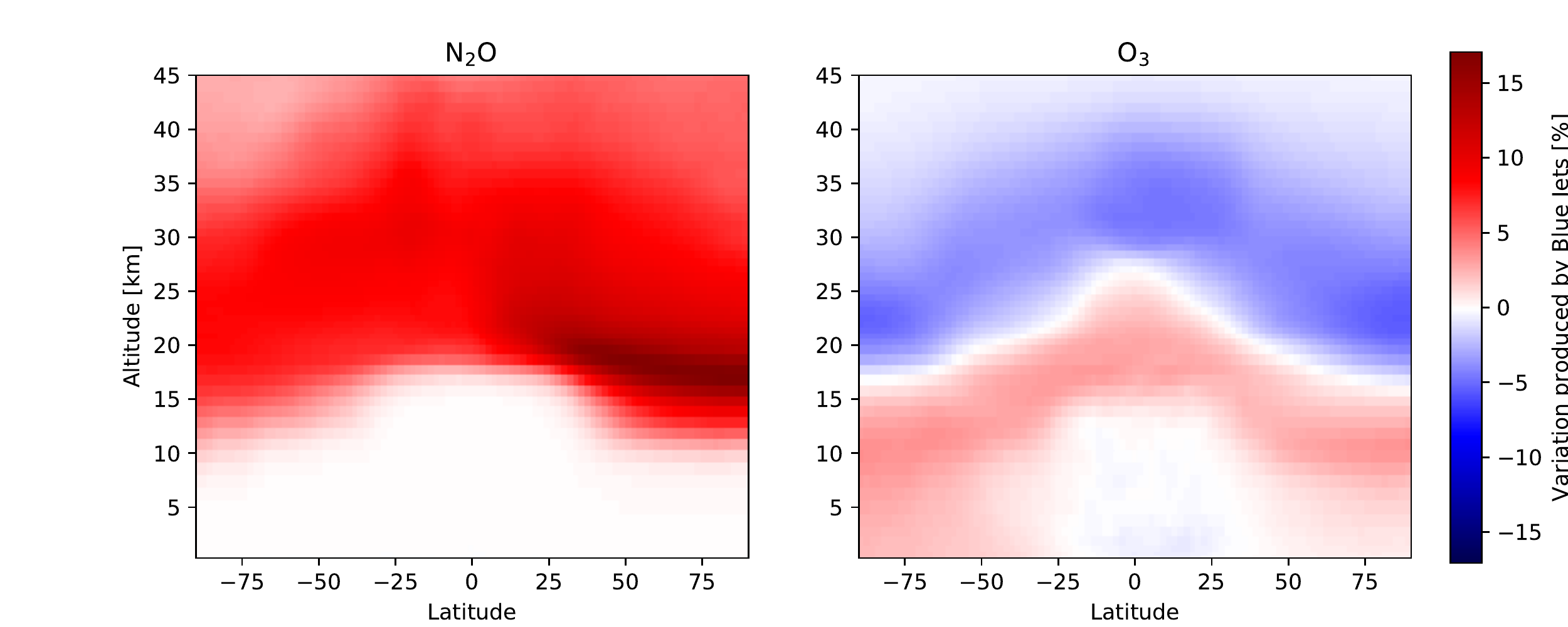}
\caption{\label{fig:lat}
Latitude-altitude distribution of the differences in the annual average density profile of N$_2$O and O$_3$ between two simulations of 10~years with (as in figure~\ref{fig:bj_column_chem_absolute}) and without Blue Jets. These variations are longitudinally average.
}
\end{figure}

Let us now investigate the geographical chemical impact of Blue Jets resulting from a 10 year simulation. We plot in figure~\ref{fig:bj_column_chem_absolute} the annual average total column density of some species after simulating a decade using the realistic BJ parameterization ``LPC-TROP LOW R$_2$".  The differences with respect to a simulation without Blue Jets are plotted in figure~\ref{fig:bj_column_chem_enhancement} (total column density) and in figure~\ref{fig:lat} (longitudinally averaged vertical profile of N$_2$O and O$_3$). The maximum influence in the density of N$_2$O and HNO$_3$ is concentrated near the North Pole, as can be seen in figures~\ref{fig:bj_column_chem_enhancement} and~\ref{fig:lat}. 
Figure~\ref{fig:lat} also shows a slight depletion of about 5 \% in the column density of O$_3$ above 30~km near the Equator. Although the total column density of O$_3$ in polar regions is not significantly affected by Blue Jets (see figure~\ref{fig:bj_column_chem_enhancement}). Figure~\ref{fig:lat} shows that there is an increase of O$_3$ below 18~km of altitude at all latitudes and a decrease above 20~km of altitude of about~5 \%. 
Some other species show differences that are distributed through mid latitudes, especially around points of maximum Blue Jet occurrence rates.

\begin{figure}
\includegraphics[width=0.6\columnwidth]{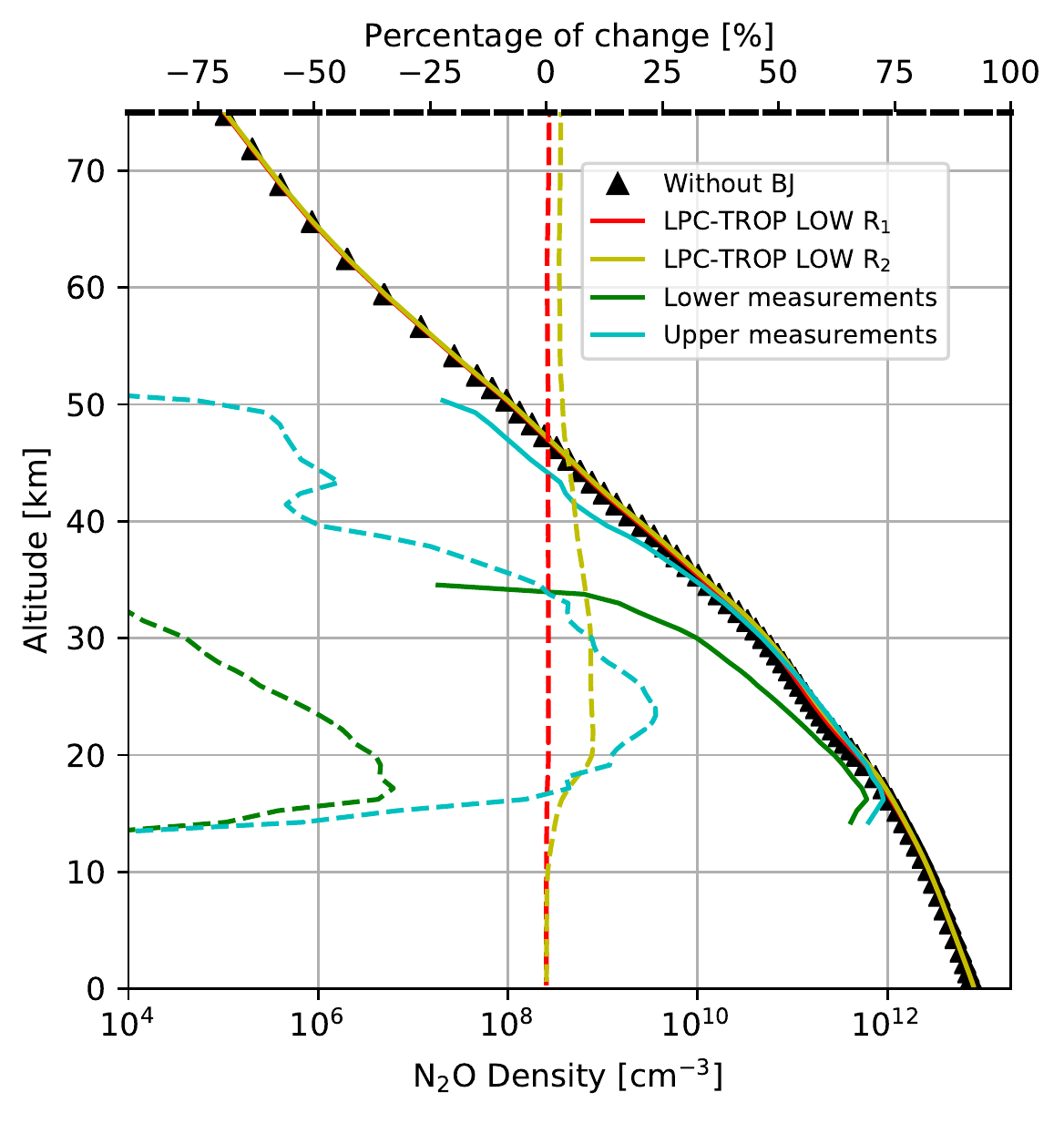}
\caption{\label{fig:bj_price_2009_N2O}
Comparison between the global average profile of N$_2$O obtained by Aura-MLS and MIPAS \citep{Plieninger2016/ACP} and the global average profile of N$_2$O after 10 year simulation with and without Blue Jets in WACCM4 (second subplot of the first row of figure~\ref{fig:bj_price_2009}) shown here with red, purple and superimposed green solid lines (absolute values) and dashed lines (percentage of change). Blue and yellow solid lines correspond to the lower and upper total N$_2$O density reported by Aura-MLS and MIPAS according to the error bars shown in figure 7 of \cite{Plieninger2016/ACP}. Blue and yellow dashed lines are the percentage of difference between the lower and upper total N$_2$O density reported by Aura-MLS and MIPAS and the N$_2$O profile of a WACCM4 simulation without Blue Jets.
}
\end{figure}

The obtained density profile of N$_2$O can be compared with measurements of Aura-MLS and MIPAS \citep{Plieninger2016/ACP} to determine how well the obtained response corresponds with observations. As detailed by \cite{Plieninger2016/ACP}, Aura-MLS and MIPAS (operating at reduced resolution mode) measured the density profile of N$_2$O over a wide range of latitudes. Figure~7 of \cite{Plieninger2016/ACP} shows the reported N$_2$O and their error bars from Aura-MLS and MIPAS. We plot these profiles together with the N$_2$O obtained after 10 year simulations with Blue Jets in figure~\ref{fig:bj_price_2009_N2O}. It can be seen that the equilibrium response of WACCM4 including Blue Jets produce a global average N$_2$O profile that falls in the range reported by Aura-MLS and MIPAS for the cases ``LPC-TROP LOW R$_1$ and R$_2$".

\normalsize

\section{Summary and conclusions}

We have introduced for the first time Blue Jets in an atmospheric global circulation model. The Blue Jet parameterization presented here is a step further in the coupling between local and global models of atmospheric electricity phenomena. 
Previous local models of Blue Jets predicted an important local enhancement of N$_2$O and NO$_x$ molecules between 18~km and 38~km of altitude, as well as a depletion of O$_3$ \citep{Winkler2015/JASTP}. The significant local chemical influence of Blue Jets suggests that their global chemical influence could be non-negligible. 

In this work, we have developed two different global parameterizations of Blue Jets. The first parameterizations (IS-TROP LOW and UP) is based on the ratio between lightning and TLE occurrence rate as reported by ISUAL and introduces a physical-based geographical dependence for the occurrence of Blue Jets. These parameterizations link the occurrence of TLEs with the altitude of the cloud top. It imposes the condition that the top of the thunderclouds must be near the tropopause in order to favor the inception of Blue Jets. Finally, the second Blue Jet parameterizations (LCP-TROP LOW and UP) are based on the observational evidences pointing to a close relationship between strong lightning discharges and Blue Jets in thunderstorms. We have obtained a good agreement between the TLE occurrence rate reported by ISUAL and the predicted ones by the Blue Jet parameterizations introduced in WACCM4 except with LCP-TROP UP. 

The implementation of these Blue Jet parameterizations in WACCM4 has allowed us to estimate their global chemical influence in the atmosphere. We have made several assumptions about the geometry of single Blue Jets in order to couple the local chemical model of \cite{Winkler2015/JASTP} with the global chemistry implemented in WACCM4. Depending on the differences between the obtained N$_2$O profile and the profiles reported by Aura-MLS and MIPAS \citep{Plieninger2016/ACP}, we have distinguished between realistic and extreme cases. According to the most realistic cases, Blue Jets would inject between 6.6 $\times$ 10$^{-4}$~Tg~N$_2$O-N~yr$^{-1}$ and 7.6~Tg~N$_2$O-N~yr$^{-1}$ near 20~km of altitude. The average value 3.8~Tg~N$_2$O-N~yr$^{-1}$ corresponds about 38~\% of natural N$_2$O sources. In addition, we have obtained that the global production of NO$_x$ by Blue Jets is between 10$^{-5}$~Tg~NO-N~yr$^{-1}$ and 0.14~Tg~NO-N~yr$^{-1}$. The average value 0.07~Tg~NO-N~yr$^{-1}$ is about 1~\% of natural NO sources, two orders of magnitude below the production of NO$_x$ by lightning on the troposphere. WACCM4 has allowed us to estimate the influence of Blue Jets in other chemical species apart from NO$_x$ and N$_2$O. In particular, we have found that the stratospheric (between 20~km and 40~km) concentration of some species such as OH, HO$_2$, SO$_2$ and HNO$_3$ could also be influenced by Blue Jets. Finally, we have also found that the inclusion of Blue Jets in WACCM4 can account for a maximum decrease of O$_3$ by about 5 \% between 20~km and 40~km of altitude.

There are several reasons behind the high uncertainty in what we call realistic results. First, there is not a clear convenient global parameterization of lightning to be combined with the proposed Blue Jet parameterizations \citep{Tost2007/ACP}. Second, the detailed mechanisms behind the production of Blue Jets are still poorly described, which makes it difficult to build global parameterization for Blue Jets. Finally, the complex chemistry taking place in the high temperature leader-phase of Blue Jets together with the uncertainties in their electrodynamical radius imply an important uncertainty of the local chemical influence of Blue Jets. All in all, we consider this work as a first approximation to the understanding of the influence of Blue Jets on the global atmospheric chemistry.

\clearpage

\small

\begin{longtable}{|c|c|c|c|c|}

\hline \multicolumn{1}{|c}{\textbf{L-BJ parameterizations}} & \multicolumn{1}{|c}{\textbf{BJ frequency [min$^{-1}$]}} & \multicolumn{1}{|c}{\textbf{Tg~NO-N~yr$^{-1}$}} & \multicolumn{1}{|c}{\textbf{Tg~N$_2$O-N~yr$^{-1}$}}  & \multicolumn{1}{|c|}{\textbf{Tg~O~yr$^{-1}$}} \\

\endfirsthead

\multicolumn{1}{|c}{\textbf{L-BJ parameterization}} & \multicolumn{1}{|c}{\textbf{BJ frequency [min$^{-1}$]}} & \multicolumn{1}{|c}{\textbf{Tg~NO-N~yr$^{-1}$}} & \multicolumn{1}{|c}{\textbf{Tg~N$_2$O-N~yr$^{-1}$}}  & \multicolumn{1}{|c|}{\textbf{Tg~O~yr$^{-1}$}} \\
\hline
\endhead

\hline

CTH IS-TROP UP  R$_1$  & 0.9  & 6  $\times$ 10$^{-3}$ & 0.42  & 3 $\times$ 10$^{-7}$ \\ 
\hline
CTH IS-TROP UP  R$_2$  & 0.9  & 0.16 & 10.4 & 7 $\times$ 10$^{-6}$ \\ 
\hline
CTH IS-TROP LOW R$_1$  & 9 $\times$ 10$^{-3}$  & 6 $\times$ 10$^{-6}$ & 4  $\times$ 10$^{-3}$ & 3 $\times$ 10$^{-9}$ \\ 
\hline
CTH IS-TROP LOW R$_2$  & 9 $\times$ 10$^{-3}$  & 1.5 $\times$ 10$^{-3}$ & 0.1 & 8 $\times$ 10$^{-8}$ \\ 
\hline
CTH LPC-TROP UP  R$_1$  & 7.2  &  5 $\times$ 10$^{-2}$ & 3.0 & 3 $\times$ 10$^{-6}$ \\ 
\hline
CTH LPC-TROP UP  R$_2$  & 7.2  & 1.36 & 76.0 & 7 $\times$ 10$^{-5}$ \\ 
\hline
CTH LPC-TROP LOW R$_1$  & 0.72  & 5 $\times$ 10$^{-3}$ & 0.3 & 3 $\times$ 10$^{-7}$ \\ 
\hline
CTH LPC-TROP LOW R$_2$  & 0.72  & 0.14 & 7.6 & 7 $\times$ 10$^{-6}$ \\ 
\hline

\caption{BJ frequency and production of NO, N$_2$O and O obtained for different one year simulations using BJ parameterizations and the CTH lightning parameterization.} \label{tab:results} \\

\end{longtable}

\section*{Acknowledgement}

The authors acknowledge helpful discussions with Rolando Garcia, Daniel Marsh, Michael Mills, Charles Bardeen, Douglas Kinnison, Andrew Gettelman, Simone Tilmes, Louisa Emmons and Heidi Huntrieser. This work was supported by the Spanish Ministry of Science and Innovation, MINECO under projects ESP2015-69909-C5-2-R and ESP2017-86263-C4-4-R and by the EU through the H2020 Science and Innovation with Thunderstorms (SAINT) project (Ref. 722337) and the FEDER program. The National Center for Atmospheric Research is sponsored by the National Science Foundation. FJPI acknowledges a PhD research contract, code BES-2014-069567. FJGV acknowledges support from the Spanish Ministry of Education and Culture under the Salvador de Madariaga program PRX17/00078. Data and codes presented here are available from figshare repository at bit.ly/WACCMBJ.

\newcommand{\pra}{Phys. Rev. A} 
\newcommand{\jgr}{J. Geoph. Res. } 
\newcommand{\jcp}{J. Chem. Phys. } 
\newcommand{\ssr}{Space Sci. Rev.} 
\newcommand{\planss}{Plan. Spac. Sci.} 
\newcommand{\pre}{Phys. Rev. E} 
\newcommand{\nat}{Nature} 
\newcommand{\icarus}{Icarus} 
\newcommand{\ndash}{-}

\end{document}